\documentclass[12pt]{article}
\pdfoutput=1
\usepackage{draft}

\usepackage{multirow}

\begin{document}

\begin{titlepage}

\begin{center}

\hfill \\
\hfill \\
\vskip 1cm

\title{
Missing Corner in the Sky: \\  Massless Three-Point Celestial Amplitudes
}

\author{Chi-Ming Chang$^{a,b}$, Wen-Jie Ma$^{b,a}$}

\address{\small${}^a$Yau Mathematical Sciences Center (YMSC), Tsinghua University, Beijing, 100084, China}

\address{\small${}^b$Beijing Institute of Mathematical Sciences and Applications (BIMSA), Beijing, 101408, China}

\email{{{\tt cmchang@tsinghua.edu.cn}, \tt wenjia.ma@bimsa.cn}}

\end{center}

\vfill

\begin{abstract}
    We present the first computation of three-point celestial amplitudes in Minkowski space of massless scalars, photons, gluons, and gravitons. Such amplitudes were previously considered to be zero in the literature because the corresponding scattering amplitudes in the plane wave basis vanish for finite momenta due to momentum conservation. However, the delta function for the momentum conservation has support in the soft and colinear regions, and contributes to the Mellin and shadow integrals that give non-zero celestial amplitudes.  We further show that when expanding in the (shadow) conformal basis for the incoming (outgoing) particle wave functions, the amplitudes take the standard form of correlators in two-dimensional conformal field theory. In particular, 
    the three-point celestial gluon amplitudes take the form of a three-point function of a spin-one current with two spin-one primary operators, which strongly supports the relation between soft spinning particles and conserved currents. Moreover, the three-point celestial amplitudes of one graviton and two massless scalars take the form of a correlation function involving a primary operator of conformal weight one and spin two, whose level-one descendent is the supertranslation current.

\end{abstract}

\vfill

\end{titlepage}

\tableofcontents


\section{Introduction}


Scattering amplitudes presented in the conformal primary basis in four-dimensional Minkowski space are naturally viewed as correlators on the celestial two-sphere at null infinity, and referred to as the celestial amplitudes \cite{Pasterski:2016qvg,Strominger:2017zoo,Raclariu:2021zjz, Pasterski:2021rjz, Pasterski:2021raf
}.
The connection between four and two dimensions provides a concrete dictionary for celestial holography, which aims to reformulate the flat space quantum gravity as a celestial conformal field theory (CCFT). The viewpoint from the celestial sphere has brought us a lot of new insights. For instance, under the celestial holography dictionary, massless spinning particles in the soft (or conformally soft) limit are dual to currents that generate infinite-dimensional symmetries in the CCFT, and the corresponding soft theorems are recast as Ward identities \cite{Weinberg:1965nx, He:2014laa, Kapec:2016jld, Donnay:2018neh,Stieberger:2018onx, Fan:2019emx, Pate:2019mfs, Adamo:2019ipt, Puhm:2019zbl, Guevara:2019ypd, Fotopoulos:2019tpe,Strominger:2021lvk, Ball:2021tmb, Strominger:2021mtt}.

From the analogy with two-dimensional conformal field theory (CFT), one expects the structure constants and central charges of this infinite-dimensional algebra are contained in the three-point correlators of the currents. However, three-point scattering amplitudes of massless particles are commonly regarded to vanish, because in the plane wave basis the momentum conservation cannot be satisfied for generic momentum. This problem is usually avoided in the literature by going to the Klein space with the unphysical $(2,2)$ split signature \cite{Pasterski:2017ylz,Stieberger:2018edy,Fan:2019emx,Pate:2019mfs,Puhm:2019zbl,Pate:2019lpp,Banerjee:2020kaa,Jiang:2021xzy,Brandhuber:2021nez,Sharma:2021gcz}. Even so, the three-point celestial gluon and graviton amplitudes still do not take the standard form of the current three-point functions in 2d CFT \cite{Pasterski:2017ylz,Puhm:2019zbl}.

 In this paper, we overcome this problem and present the first computation of three-point scattering amplitudes of massless particles in the Lorentzian $(-+++)$ signature. The computation relies on an important observation that solutions to the total momentum conservation of three massless particles in Minkowski space do exist in two special regions, and give nontrivial contributions to the massless three-point celestial amplitudes. They are the soft region where one of the particles has zero energy, and the colinear region where the momenta of all the incoming and outgoing particles are colinear. The three-point celestial amplitudes with only massless scalars receive contributions from both limits regions (see Section \ref{sec:scalar_amplitudes}). 
The three-point celestial amplitudes involving massless spinning particles vanish in the colinear region, due to the tensor structure in the cubic vertex. Non-zero amplitudes in the (conformally) soft region are extracted using the logarithmic conformal primary wave functions obtained in \cite{Donnay:2018neh} (see Section \ref{sec:l-0-0_amplitudes} and \ref{sec:l-l-l_amplitudes}). 

The results are best presented when the incoming (outgoing) particles are in the (shadow) conformal primary basis. In this prescription, scattering amplitudes are given by the shadow products between the incoming and outgoing states, which correspond to the Belavin-Polyakov-Zamolodchikov (BPZ) inner products in the CCFT \cite{Crawley:2021ivb}. The celestial amplitudes in this prescription are summarized in Table \ref{tab:celestial_amplitudes}. We can see that all of them take the standard form of the correlators of primary operators in 2d CFT. In particular, the soft photon and gluon correspond to the spin-1 conserved currents, and the soft graviton corresponds to a primary operator of dimension $(\frac{3}{2},-\frac{1}{2})$, whose level-1 descendent is the supertranslation current \cite{Donnay:2018neh}.


\begin{table}[H]
\begin{center}
\begin{tabular}{c|c|c}
particles & celestial amplitudes & operator dimensions
\\ \hline && \\[-0.45cm]
scalar-photon-scalar & \multicolumn{1}{l|}{$\,\,\mathcal{A}_{1^0\to 2_{\rm soft}^-3^0}\propto \left(\frac{1}{z_{12}}+\frac{1}{z_{23}}\right)\frac{1}{|z_{13}|^{2\Delta}}$} & $(\frac{\Delta}{2},\frac{\Delta}{2})$, $(1,0)$, $(\frac{\Delta}{2},\frac{\Delta}{2})$
\\[0.2cm] \hline && \\[-0.45cm]
scalar-graviton-scalar &  \multicolumn{1}{l|}{$\,\,\mathcal{A}_{1^0\rightarrow2^{--}_{\rm soft}3^0}
\propto\frac{\bar z_{12}z_{13}^2}{z_{12} z_{23}^2}\frac{1}{|z_{13}|^{2\Delta+2}}$}  & $(\frac{\Delta}{2},\frac{\Delta}{2})$, $(\frac{3}{2},-\frac{1}{2})$, $(\frac{\Delta+1}{2},\frac{\Delta+1}{2})$
\\[0.2cm] \hline && \\[-0.45cm]
\multirow{3}{*}{three gluons} & $\mathcal{A}_{1^+\to 2^-_{\rm soft}3^-}\propto\left(\frac{1}{z_{12}}+\frac{1}{z_{23}}\right)\frac{1}{z_{13}^{\Delta+1}\bar{z}_{13}^{\Delta-1}}$ & $(\frac{\Delta+1}{2},\frac{\Delta-1}{2})$, $(1,0)$, $(\frac{\Delta+1}{2},\frac{\Delta-1}{2})$
\\[0.2cm] \cline{2-3} && \\[-0.45cm]
&  $\mathcal{A}_{1^-\to 2^-_{\rm soft}3^+}\propto\left(\frac{1}{z_{12}}+\frac{1}{z_{23}}\right)\frac{1}{z_{13}^{\Delta-1}\bar{z}_{13}^{\Delta+1}}$ & $(\frac{\Delta-1}{2},\frac{\Delta+1}{2})$, $(1,0)$, $(\frac{\Delta-1}{2},\frac{\Delta+1}{2})$
\end{tabular}
\end{center}
\caption{\label{tab:celestial_amplitudes}The celestial amplitudes of two massless scalars with a photon or a graviton, and the three-point celestial gluon amplitudes.  The logarithmic conformal primary wave functions were used for the soft (second) particles. The conformal dimensions of the corresponding CCFT operators are given in the last column.}
\end{table}

The remainder of this paper is organized as follows. Section \ref{sec:ConformalBasis} reviews the celestial amplitude and the conformal primary basis of low spin, gives a general formula for massless and massive spin-$\ell$ conformal primary basis, and discusses the massless limit of the massive conformal primary basis. Section \ref{sec:scalar_amplitudes} computes the three-point celestial amplitudes of massless scalars. Section \ref{sec:l-0-0_amplitudes} computes the celestial amplitudes of two massless scalars with a photon or a graviton. Section \ref{sec:l-l-l_amplitudes} computes the three-point celestial gluon amplitudes. Section \ref{sec:discussion} ends with a summary and future directions.



\section{Conformal primary basis}\label{sec:ConformalBasis}

Celestial amplitudes are obtained by expanding the position space amplitudes with respect to the conformal primary wavefunctions \cite{Pasterski:2016qvg,Pasterski:2017kqt} instead of the plane-waves, \textit{i.e.,} \footnote{Throughout this paper, $\mathcal{O}_i(z_i)$ should be understood as $\mathcal{O}_i(z_i,\bar{z}_i)$. We use this abbreviation to simplify the notation.}
\begin{align}\label{eq:CA}
\mathcal{A}^{\Delta_i}_{a_i}(z_i)=\bigg(\prod_{j=1}^{k+n}\int d^{4}X_j\bigg)\bigg(\prod_{j=1}^k\phi_{\Delta_j,a_j}^{+}(z_j;X_j)\bigg)\bigg(\prod_{j=k+1}^{k+n}\widetilde{\phi}^{-}_{\Delta_j,a_j}(z_j;X_{j})\bigg)\mathcal{M}(X_i)\;,
\end{align}
where $\mathcal{M}(X_j)$ is the scattering amplitude in position space, and $\phi^{\pm}_{\Delta,a}(z;X)$, $\widetilde\phi^{\pm}_{\Delta,a}(z;X)$ are the conformal primary and shadow conformal primary wave functions. $\Delta$ and $a$ are the conformal dimension and the spin of the wave functions, respectively. The coordinates $(z,\bar z)$ on the celestial sphere are related to the massless on-shell momenta $q^\mu$ through
\begin{align}\label{hq}
q^{\mu}(\omega,z)=\omega\hat{q}^{\mu}(z)=\omega(1+z\bar{z},z+\bar{z},-i(z-\bar{z}),1-z\bar{z})
\end{align}
and to the massive on-shell momenta $p^{\mu}$ through
\begin{align}\label{hp}
p^{\mu}(m,y,z)= m\hat{p}^{\mu}(y,z)=\frac{m}{2y}(1+y^2+z\bar{z},z+\bar{z},-i(z-\bar{z}),1-y^2-z\bar{z})\,.
\end{align}
Here, we choose to expand the wave functions of the incoming particles in the conformal primary basis while the outgoing particles in the shadow conformal primary basis. This prescription is equivalent to expanding the wave functions of both incoming and outgoing particles in the same conformal primary basis but modifying the inner product in the definition of the $S$-matrix from the Klein-Gordon inner product to the shadow product \cite{Crawley:2021ivb}. Finally, note that while the helicity in four dimensions aligns with the spin on the celestial sphere for conformal primary basis $\phi^{\pm}_{\Delta,a}$, the helicity and spin have opposite signs for the shadow conformal primary basis $\widetilde\phi^{\pm}_{\Delta,a}$ due to the shadow transform.

\subsection{Scalar}

The conformal primary wave functions $\phi^{\pm}_{\Delta}(z;X)$ for massless and massive scalars with mass $m$ are given by
\begin{align}\label{eq:Massless}
\phi^{\pm}_{\Delta}(z;X)= \int_{0}^{+\infty}d\omega\,\omega^{\Delta-1}e^{\pm i\omega\hat{q}\cdot X-\epsilon\omega}=\frac{(\mp i)^{\Delta}\Gamma[\Delta]}{(-\hat{q}\cdot X\mp i\epsilon)^{\Delta}}\equiv N^{\pm}_\Delta\frac{1}{(-\hat{q}\cdot X\mp i\epsilon)^{\Delta}}\;,
\end{align}
and
\begin{align}\label{eq:Massive}
\phi^{\pm}_{\Delta,m}(z;X)=\int\frac{d^3\hat{p}^{\prime}}{\hat{p}^{\prime0}}G_{\Delta}(z,\bar{z};\hat{p}^{\prime})e^{\pm im\hat{p}^{\prime}\cdot X}\;,
\end{align}
respectively. Here, $G_{\Delta}(z,\bar{z};\hat{p})$ is the bulk-to-boundary propagator which takes the form as
\begin{align}\label{eqn:bdry_to_bulk}
G_{\Delta}(\hat{q};\hat{p}^{\prime})=\frac{1}{(-\hat{q}\cdot\hat{p}^{\prime})^{\Delta}}
\end{align}
in terms of $\hat q$ and $\hat{p}$ in \eqref{hq} and \eqref{hp}. \footnote{In this paper, we use the most positive metric in four-dimensional flat space, \textit{i.e.}, $\eta_{\mu\nu}=\text{diag}(-1,+1,+1,+1)$.} Starting with \eqref{eq:Massless} and \eqref{eq:Massive}, one can obtain another set of conformal primary basis by performing the shadow transformations. As shown in \cite{Pasterski:2017kqt}, the shadow transformation of massive conformal primary basis \eqref{eq:Massive} takes the same form as \eqref{eq:Massive} up to a change of conformal dimension from $\Delta$ to $2-\Delta$. On the other hand, the shadow transformation of massless conformal primary basis \eqref{eq:Massless} is 
\begin{align}\label{eq:phit}
    \widetilde{\phi}_{\Delta}^{\pm}(z;X)=2^{\Delta-1}\frac{\Gamma[2-\Delta]}{2\pi\Gamma[1-\Delta]}\int\frac{d^{3}q^{\prime}}{q^{0\prime}}\frac{1}{(-q^{\prime}\cdot\hat{q})^{\Delta}}e^{\pm iq^{\prime}\cdot X}=\frac{\Gamma[2-\Delta]}{2\pi\Gamma[1-\Delta]}\int d^2z^{\prime}\frac{\phi_{2-\Delta}^{\pm}(z^{\prime};X)}{|z-z^{\prime}|^{2\Delta}}
\;.
\end{align}
The integral over $d^2z^{\prime}$ can be computed, leading to the following relation \cite{Pasterski:2017kqt}
\begin{align}\label{eq:phit=phi}
    \widetilde{\phi}_{\Delta}^{\pm}(z;X)=\frac{N^\pm_{2-\Delta}}{N^\pm_{\Delta}}(-X^2)^{\Delta-1}\phi_{\Delta}^{\pm}(z;X)\;.
\end{align}
%


\subsection{Massless spin-one}

The conformal primary basis for massless spin-one particles is \cite{Pasterski:2017kqt,Donnay:2018neh}
\begin{align}\label{eq:SpinOne}
\begin{split}
A^{\Delta,\pm}_{\mu a}(\hat{q};X)&=\frac{\Delta-1}{\Delta}\frac{\partial_{a}\hat{q}^{\mu}}{(-\hat{q}\cdot X\mp i\epsilon)^{\Delta}}+\partial_\mu\frac{\partial_{a}\hat{q}\cdot X}{\Delta(-\hat{q}\cdot X\mp i\epsilon)^{\Delta}}
\\
&\equiv V^{\Delta,\pm}_{\mu a}(\hat{q};X)+\partial_\mu\alpha^{\Delta,\pm}_{a}(\hat{q};X)\;.
\end{split}    
\end{align}
The corresponding shadow conformal primary wave function is related to $A^{\Delta,\pm}_{\mu a}$ by
\begin{align}\label{eq:At=A}
\begin{split}
 \widetilde{A}^{\Delta,\pm}_{\mu a}=(-X^2)^{\Delta-1}A^{\Delta,\pm}_{\mu a}\;,
\end{split}    
\end{align}
leading to
\begin{align}\label{eq:At}
\begin{split}
 \widetilde{A}^{\Delta,\pm}_{\mu a}
 =&\frac{(-X^2)^{\Delta-1}\partial_{a}\hat{q}^{\mu}}{(-\hat{q}\cdot X\mp i\epsilon)^{\Delta}}+\frac{(-X^2)^{\Delta-1}\hat{q}^{\mu}\partial_{a}\hat{q}\cdot X}{(-\hat{q}\cdot X\mp i\epsilon)^{\Delta+1}}\;.
\end{split}    
\end{align}

From \eqref{eq:SpinOne}, it is easy to check that $A^{\Delta,\pm}_{\mu a}(\hat{q};X)$ is gauge equivalent to the Mellin transformation of the plane-wave basis up to a constant when $\Delta\neq1$, \textit{i.e.} 
\begin{align}\label{eq:SpinOneMellin}
A^{\Delta,\pm}_{\mu a}(\hat{q};X)=\frac{\Delta-1}{\Delta N^{\pm}_\Delta}\partial_{a}\hat{q}^{\mu}\int_0^{\infty}d\omega\,\omega^{\Delta-1}e^{\pm i\omega\hat{q}\cdot X-\epsilon\omega}+\partial_\mu\alpha^{\Delta,\pm}_{a}\;.    
\end{align}
On the other hand, in the conformally soft region $\Delta=1$, both $A^{\Delta,\pm}_{\mu a}$ and $\widetilde{A}^{\Delta,\pm}_{\mu a}$ reduce to pure gauge and lead to vanishing celestial amplitude due to the gauge invariance. The conformal primary wave functions that are not pure gauge can be constructed from the combination
of $A^{\Delta,\pm}_{\mu a}$ and $\widetilde{A}^{\Delta,\pm}_{\mu a}$:
\begin{align}\label{eq:Alog_def}
\begin{split}
   A^{\text{log},\pm}_{\mu a}=\lim_{\Delta\rightarrow 1}\partial_{\Delta}\bigg(A^{\Delta,\pm}_{\mu a}+\widetilde{A}^{2-\Delta,\pm}_{\mu a}\bigg)\;. 
\end{split}    
\end{align}
With the help of \eqref{eq:At=A}, the log mode $A^{\text{log},\pm}_{\mu a}$ can be further written as
\begin{align}\label{eq:Alog}
\begin{split}
   A^{\text{log},\pm}_{\mu a}(\hat{q};X)&=-\log(-X^2)A^{1,\pm}_{\mu a}(\hat{q};X)
   \\
   &=\frac{2X_{\mu}}{X^2}\frac{\partial_{a}\hat{q}\cdot X}{(-\hat{q}\cdot X\mp i\epsilon)}+\partial_\mu\left(-\log(-X^2)\frac{\partial_{a}\hat{q}\cdot X}{(-\hat{q}\cdot X\mp i\epsilon)}\right)
   \\
   &\equiv V^{\text{log},\pm}_{\mu a}(\hat{q};X)+\partial_\mu\alpha^{\text{log},\pm}_{a}(\hat{q};X)\;.
\end{split}    
\end{align}

In the later computations, we will also use the wave functions in momentum space. Since $A^{\Delta,\pm}_{\mu a}$ satisfies the massless Klein-Gordon equation $\Box A^{\Delta,\pm}_{\mu a}=0$, it admits the Fourier expansion in the on-shell momentum space as
\begin{align}\label{eq:SpinOneFourier}
A^{\Delta,\pm}_{\mu a}(\hat{q};X)=&\int\frac{d^3q^{\prime}}{q^{\prime}_0}A^{\Delta,\pm}_{\mu a}(\hat{q};q^{\prime})e^{\pm iq^{\prime}\cdot X}\;,
\end{align}
where the integral is over all null momenta $q^{\prime\mu}$. The Lorentz gauge condition $\partial^{\mu}A^{\Delta,\pm}_{\mu a}(\hat{q};X)=0$ in the momentum space becomes the transverse condition  $q^{\prime\mu} A^{\Delta,\pm}_{\mu a}(\hat{q};q^{\prime})=0$. Similarly, one has the Fourier modes $\alpha^{\Delta,\pm}_{a}(\hat{q};q^{\prime})$, $V_{\mu a}^{\Delta,\pm}(\hat{q};q^{\prime})$,  $\widetilde{A}^{\Delta,\pm}_{\mu a}(\hat{q};q^{\prime})$ and ${A}^{\text{log},\pm}_{\mu a}(\hat{q};q^{\prime})$, which are functions of the on-shell momentum $q'^\mu$, and the latter three satisfy the transverse condition. However, $V^{\text{log},\pm}_{\mu a}$ and $\alpha^{\text{log},\pm}_a$ do not admit an expansion in the on-shell momentum space like \eqref{eq:SpinOneFourier}, because they do not satisfy the massless Klein-Gordon equation. 
After defining the Fourier modes, \eqref{eq:SpinOne} can be translated into the momentum space, giving
\begin{align}
A^{\Delta,\pm}_{\mu a}(\hat{q};q^{\prime})=& V^{\Delta,\pm}_{\mu a}(\hat{q};q^{\prime})\pm iq^{\prime}_\mu\alpha^{\Delta,\pm}_{a}(\hat{q};q^{\prime})\;.
\end{align}
%
%
%

\subsection{Massless spin-two}

The conformal primary basis for massless spin-two particles has been studied in \cite{Pasterski:2017kqt,Donnay:2018neh}. It takes the form as
\begin{align}
\begin{split}
h^{\Delta}_{\mu\nu zz}(\hat q, X)&=\frac{1}{2}\left[\frac{\partial_z \hat q_\mu \partial_z \hat q_\nu}{(-\hat q\cdot X\mp i\epsilon)^\Delta }+\frac{2(\partial_z\hat q\cdot X)\hat q_{(\mu}\partial_z \hat q_{\nu)}}{(-\hat q\cdot X\mp i\epsilon)^{\Delta+1}}+\frac{(\partial_z\hat q\cdot X)^2 \hat q_\mu \hat q_\nu}{(-\hat q\cdot X\mp i\epsilon)^{\Delta+2}}\right]\;.
\end{split}
\end{align}
It is convenient to decompose $h^{\Delta}_{\mu\nu zz}(\hat q, X)$ as
\begin{align}\label{eq:SpinTwo}
\begin{split}
h^{\Delta}_{\mu\nu zz}(\hat q, X)&=\frac{\Delta-1}{2(\Delta+1)}\frac{\partial_z \hat q_\mu \partial_z \hat q_\nu}{(-\hat q\cdot X\mp i\epsilon)^\Delta }
\\
&\quad+\frac{(\Delta-1)}{\Delta(\Delta+1)}\hat q_{(\mu}\partial_{\nu)}\frac{(\partial_z \hat q\cdot X)}{(-\hat q\cdot X\mp i\epsilon)^\Delta}+\frac{1}{2\Delta(\Delta+1)}\partial_\mu\partial_\nu\frac{(\partial_z \hat q\cdot X)^2}{(-\hat q\cdot X\mp i\epsilon)^\Delta}\;,
\end{split}
\end{align}
where the terms on the second line are total derivatives.
The corresponding shadow conformal primary wave function is related to $h^{\Delta,\pm}_{\mu\nu zz}$ by
\begin{align}\label{eq:ht=h}
\begin{split}
 \widetilde{h}^{\Delta,\pm}_{\mu\nu zz}=(-X^2)^{\Delta-1}h^{\Delta,\pm}_{\mu\nu zz}\;.
\end{split}    
\end{align}
We note that when $\Delta\neq0$ and $\Delta\neq 1$, $h^{\Delta}_{\mu\nu zz}(\hat q, X)$ is proportional to the Mellin transformation of the plane-wave basis up to pure diffeomorphism:
\begin{align}
\begin{split}
h^{\Delta}_{\mu\nu zz}(\hat q, X)&=\frac{(\Delta-1)}{2(\Delta+1)N^\pm_\Delta}\partial_z \hat q_\mu \partial_z \hat q_\nu\int_{0}^{\infty}d\omega\,\omega^{\Delta-1}e^{\pm i\omega\hat{q}\cdot X-\epsilon\omega}+(\text{total derivative})\;.
\end{split}
\end{align}
In the conformally soft limit $\Delta\rightarrow1$, the conformal primary wave function \eqref{eq:SpinTwo} and its shadow \eqref{eq:ht=h} coincide and both reduce to pure diffeomorphism. In this limit, a new conformal primary wave function $h^{\text{log},\pm}_{\mu\nu zz}$, which is not pure diffeomorphism, can be constructed from
\begin{align}\label{eq:hlog}
\begin{split}
&h^{\text{log},\pm}_{\mu\nu zz}=\lim_{\Delta\rightarrow1}\partial_{\Delta}\bigg(h^{\Delta}_{\mu\nu zz}(\hat q, X)+\widetilde{h}^{2-\Delta}_{\mu\nu zz}(\hat q, X)\bigg)\;.
\end{split}
\end{align}
Armed with \eqref{eq:SpinTwo} and \eqref{eq:ht=h}, $h^{\text{log},\pm}_{\mu\nu zz}$ can be re-written as
\begin{align}
\begin{split}
&h^{\text{log},\pm}_{\mu\nu zz}=-\text{log}(-X^2)h^{\Delta=1}_{\mu\nu zz}(\hat q, X)=W^{\text{log},\pm}_{\mu\nu zz}+\partial_{(\mu}\alpha_{\nu)}^{\text{log},\pm}\;,
\end{split}
\end{align}
where $W^{\text{log},\pm}_{\mu\nu zz}$ is 
\begin{align}\label{eq:WLog}
\begin{split}
W^{\text{log},\pm}_{\mu\nu zz}=\frac{\eta_{\mu\nu}-\frac{2X_{\mu}X_{\nu}}{X^2}}{2(-X^2)}\frac{(\partial_z \hat q\cdot X)^2}{(-\hat q\cdot X\mp i\epsilon)}\;.
\end{split}
\end{align}
and $\alpha^{\text{log},\pm}_{\mu zz}$ is just some function of $\hat{q}$ and $X$.

\subsection{General spin}\label{sec:CPB_general_spin}

In this subsection, we give a general formula for the massless and massive conformal primary basis of arbitrary integer spin, which is derived from a covariant formalism introduced in Appendix \ref{app:CPB_general_spin}. Under specialization,  one can get various conformal primary basis including those reviewed in the previous subsections, and the massive spin-$\ell$ conformal primary basis obtained \cite{Law:2020tsg}.

The spin-$\ell$ massless conformal primary basis is\footnote{We use the abbreviation ${\{\nu^{\ell}\}}\equiv {\nu_1\cdots\nu_{\ell}}$ and $(\partial_a \hat q_\nu)^\ell\equiv \partial_{a_1} \hat q_{\nu_1}\cdots \partial_{a_\ell} \hat q_{\nu_\ell}$ to simplify the notation.}
\begin{align}\label{eq:MasslessSpinl}
\begin{split}
\Phi^{\Delta,\pm}_{\{\mu^{\ell}\}\{a^{\ell}\}}(\hat{q};X)=&\mathcal{N}_{\Delta,\ell}(\partial_a\hat{q}_{\nu})^{\ell}\int\frac{d^{3}q^{\prime}}{q^{0\prime}}\frac{{(\mathcal{P}_I^{\ell})_{\{\mu^{\ell}\}}}^{\{\nu^{\ell}\}}(\hat{q},\hat{q}^{\prime})}{(-q^{\prime}\cdot\hat{q})^{\Delta}}e^{\pm iq^{\prime}\cdot X}\;,
\end{split}
\end{align}
where $\mathcal{N}_{\Delta,\ell}$ is a normalization constant
\begin{align}\label{eq:NDelta}
\mathcal{N}_{\Delta,\ell}=\frac{\Gamma[2-\Delta+\ell]}{2\pi\Gamma[1-\Delta+\ell]}\;,  
\end{align}
and $\mathcal{P}_I^{\ell}$ is the spin-$\ell$ projection operator of $SO(2)$
\begin{align}\label{eq:PI}
\begin{split}
(\mathcal{P}_I^{\ell})_{\mu_1\cdots \mu_{\ell}}^{\phantom{\mu_1\cdots \mu_{\ell}}\nu_1\cdots \nu_\ell}(\hat{q},\hat{q}^{\prime})=&\sum_{i=0}^{\lfloor\ell/2\rfloor}\frac{(-\ell)_{2i}}{2^{2i}i!(-\ell+1)_i}I_{(\mu_1\mu_2}I^{(\nu_1\nu_2}\cdots I_{\mu_{2i-1}\mu_{2i}}I^{\nu_{2i-1}\nu_{2i}}\\
&\phantom{=}\qquad\qquad\times I_{\mu_{2i+1}}^{\phantom{\mu_{2i+1}}\nu_{2i+1}}\cdots I_{\mu_{\ell})}^{\phantom{\mu_{\ell}}\nu_{\ell})}\;,
\\
I_{\mu\nu}(\hat{q},\hat{q}^{\prime})=&\eta_{\mu\nu}-\frac{\hat{q}_{\mu}\hat{q}^{\prime}_{\nu}}{\hat{q}\cdot\hat{q}^{\prime}}-\frac{\hat{q}_{\nu}\hat{q}^{\prime}_{\mu}}{\hat{q}\cdot\hat{q}^{\prime}}\;.
\end{split}
\end{align}
In Appendix \ref{app:SCPB}, we show that the specialization of \eqref{eq:PI} to $\ell=0$ and $1$ are proportional to the shadow conformal primary basis $\widetilde\phi^\pm_\Delta$ and $\widetilde{A}^{\Delta,\pm}_{\mu a}$ as
\begin{align}
\begin{split}\label{eq:MasslessScalarShadow}
\Phi^{\Delta,\pm}(\hat{q};X)=&2^{1-\Delta}\widetilde\phi^{\pm}_{\Delta}(z;X)\;,
\\
\Phi^{\Delta,\pm}_{\mu a}(\hat{q};X)=&2^{1-\Delta}(\mp i)^{2-\Delta}\frac{\Gamma[3-\Delta]}{1-\Delta}\widetilde{A}^{\Delta,\pm}_{\mu a}(\hat{q};X)\;.
\end{split}
\end{align}

The spin-$\ell$ massive conformal primary basis is
\begin{align}\label{eq:MassiveSpin}
\begin{split}
 \Phi^{\Delta,\ell,J,\pm}_{\{\mu^{\ell}\}\{a^{|J|}\};m}(\hat{q};X)=\mathcal{N}_{\Delta,\ell}\int\frac{d^3p}{p^0}\frac{(-1)^{\ell-|J|}(\mathcal{P}^{\ell}_{P})_{\mu^{\ell}}^{\phantom{\mu^{\ell}}\mu^{\prime\ell}}(\hat{q}_{\mu^{\prime}})^{\ell-|J|}(\mathcal{P}^{|J|}_{Q})_{\mu^{\prime|J|}}^{\phantom{\mu^{\prime|J|}}\nu^{|J|}}(\partial_a\hat{q}_{\nu})^{|J|}}{(-\hat{q}\cdot p)^{\Delta+\ell-|J|}}e^{\pm ip\cdot X}\;. 
\end{split}    
\end{align}
where $J=-\ell,\,-\ell+1,\,\cdots,\,\ell$ is the eigenvalue of the Cartan generator of the $SO(3)$ little group and corresponds to the spin on the celestial sphere. The tensor structures ${\cal P}^\ell_P$ and ${\cal P}^\ell_Q$ are given by \eqref{eq:PI} with $I$ replaced by the projectors
\begin{align}
\begin{split}
&P^{\mu\nu}=\eta^{\mu\nu}+\hat{p}^{\mu}\hat{p}^{\nu}\;,\\
&Q^{\mu\nu}=\eta^{\mu\nu}-\frac{\hat{p}^{\mu}\hat{q}^{\nu}}{\hat{p}\cdot\hat{q}}-\frac{\hat{p}^{\nu}\hat{q}^{\mu}}{\hat{p}\cdot\hat{q}}-\frac{\hat{q}^{\mu}\hat{q}^{\nu}}{(\hat{p}\cdot\hat{q})^2}.
\end{split}
\end{align}
For the $J=\ell=0$ case, \eqref{eq:MassiveSpin} coincides with  \eqref{eq:Massive} up to a constant factor
\begin{align}\label{eqn:massive_scalar}
\begin{split}
\Phi^{\Delta,\pm}_m(\hat q;X)=\frac{m^{2-\Delta}(1-\Delta)}{2\pi}\phi^\pm_{\Delta,m}(z;X)\,.
\end{split}    
\end{align}
%

\subsection{Massless limit}

One advantage of \eqref{eq:MassiveSpin} is that it enjoys a nice behaviour under massless limit. Particularly, \eqref{eq:MassiveSpin} with $|J|=\ell$ directly reduces to the corresponding massless spin-$\ell$ conformal primary basis in the massless limit $m\rightarrow0$. To see this, we note that setting $J=\ell$ in \eqref{eq:MassiveSpin} leads to
\begin{align}
\begin{split}
 \Phi^{\Delta,\ell,\ell,\pm}_{\{\mu^{\ell}\}\{z^{\ell}\};m}(\hat{q};X)=\mathcal{N}_{\Delta,\ell}\int\frac{d^3p}{p^0}\frac{(\mathcal{P}^{\ell}_{P})_{\{\mu^{\ell}\}}^{\phantom{\mu^{\ell}}\{\mu^{\prime\ell}\}}(\mathcal{P}^{\ell}_{Q})_{\{\mu^{\prime\ell}\}}^{\phantom{\mu^{\prime\ell}}\{\nu^{\ell}\}}(\partial_z\hat{q}_{\nu})^{\ell}}{(-\hat{q}\cdot p)^{\Delta}}e^{\pm ip\cdot X}\;. 
\end{split}    
\end{align}
Using the transversality condition \eqref{eq:TransC} and the fact that $\mathcal{P}^{\ell}_{Q}$ is traceless with respect to the metric $g$, we can remove the projection operator $\mathcal{P}^{\ell}_{P}$ and get
\begin{align}
\begin{split}
 \Phi^{\Delta,\ell,\ell,\pm}_{\{\mu^{\ell}\}\{z^{\ell}\};m}(\hat{q};X)=\mathcal{N}_{\Delta,\ell}\int\frac{d^3p}{p^0}\frac{(\mathcal{P}^{\ell}_{Q})_{\mu^{\ell}}^{\phantom{\mu^{\ell}}\nu^{\ell}}(\partial_z\hat{q}_{\nu})^{\ell}}{(-\hat{q}\cdot p)^{\Delta}}e^{\pm ip\cdot X}\;. 
\end{split}    
\end{align}
After defining $\omega=\frac{m}{2y}$ and $\bar{p}=(1+y^2+z^{\prime}\bar{z}^{\prime},z^{\prime}+\bar{z}^{\prime},-i(z^{\prime}-\bar{z}^{\prime}),1-y^2-z^{\prime}\bar{z}^{\prime})$, we have $p=m\hat{p}=\omega\bar{p}$, leading to
\begin{align}\label{eq:MasslessLimit}
\begin{split}
&\int\frac{d^3p}{p^0}\frac{1}{(-\hat{q}\cdot p)^{\Delta}}=2^{1-\Delta}\int_0^{\infty}d\omega\,\omega^{1-\Delta}\int d^2z^{\prime}\bigg(\frac{1}{\frac{m^2}{4\omega^2}+|z-z^{\prime}|^2}\bigg)^{\Delta}\;,
\end{split}
\end{align}
and
\begin{align}\label{eq:MasslessLimit1}
\begin{split}
&Q^{\mu\nu}=\eta^{\mu\nu}-\frac{\bar{p}^{\mu}\hat{q}^{\nu}}{\bar{p}\cdot\hat{q}}-\frac{\bar{p}^{\nu}\hat{q}^{\mu}}{\bar{p}\cdot\hat{q}}-\frac{m^2}{\omega^2}\frac{\hat{q}^{\mu}\hat{q}^{\nu}}{(\bar{p}\cdot\hat{q})^2}\;.
\end{split}
\end{align}
Thus, in the massless limit $m\rightarrow 0$, we get
\begin{align}\label{eq:MasslessLimitGeneric}
\begin{split}
\lim_{m\rightarrow0}\Phi^{\Delta,\ell,\ell,\pm}_{\{\mu^{\ell}\}\{z^{\ell}\};m}(\hat{q};X)=&2^{1-\Delta}\mathcal{N}_{\Delta,\ell}\int_0^{\infty}d\omega\,\omega^{1-\Delta}\int d^2z^{\prime}\frac{(\mathcal{P}^{\ell}_{I})_{\mu^{\ell}}^{\phantom{\mu^{\ell}}\nu^{\ell}}(\partial_z\hat{q}_{\nu})^{\ell}}{|z-z^{\prime}|^{2\Delta}}e^{\pm iq^{\prime}\cdot X}\\
 =&2^{1-\Delta}\mathcal{N}_{\Delta,\ell}\int\frac{d^3q^{\prime}}{q^{\prime0}}\frac{(\mathcal{P}^{\ell}_{I})_{\mu^{\ell}}^{\phantom{\mu^{\ell}}\nu^{\ell}}(\partial_z\hat{q}_{\nu})^{\ell}}{(-\hat{q}\cdot q^{\prime})^{\Delta}}e^{\pm iq^{\prime}\cdot X}\\
 =&\Phi^{\Delta,\ell,\pm}_{\{\mu^{\ell}\}\{z^{\ell}\}}(\hat{q};X)\;,
\end{split}    
\end{align}
where we used the fact that
\begin{align}
\lim_{m\rightarrow 0}\bar{p}=(1+z^{\prime}\bar{z}^{\prime},z^{\prime}+\bar{z}^{\prime},-i(z^{\prime}-\bar{z}^{\prime}),1-z^{\prime}\bar{z}^{\prime})\equiv\frac{1}{\omega}q^{\prime}\;. 
\end{align}

For $\ell=0$, there is a different massless limit
\begin{align}\label{eq:MasslessLimitScalar1}
\begin{split}
\lim_{m\rightarrow0}m^{2\Delta-2}\Phi^{\Delta,\pm}_{m}(\hat{q};X)=-2^{\Delta-1}\phi^{\pm}_{\Delta}(z;X)\;,
\end{split}
\end{align}
which produces the massless scalar conformal primary wave function \eqref{eq:Massless}. This formula follows directly from the limit
\begin{align}\label{eq:mLimit=delta_0}
\begin{split}
\lim_{m\rightarrow0}(\Delta-1)m^{2\Delta-2}\bigg(\frac{1}{\frac{m^2}{4\omega^2}+|z-z^{\prime}|^2}\bigg)^{\Delta}=2^{2\Delta-1}\pi\omega^{2\Delta-2}\delta^{(2)}(z-z^{\prime})
\end{split}
\end{align}
which holds when $\text{Re}(\Delta)\geq1$ and can be proved as follows. We note that the massless limit $m\rightarrow0$ in \eqref{eq:mLimit=delta_0} vanishes unless $z=z^{\prime}$. Thus the left-hand side of \eqref{eq:mLimit=delta_0} must be proportional to $\delta^{(2)}(z-z^{\prime})$. \footnote{One may expect that when $\omega=0$ the massless limit is also non-vanishing. However, due to the fact that $\text{Re}(2-2\Delta)\geq0$, we have
\begin{align}
\begin{split}
\frac{2^{\Delta-2}(1-\Delta)}{\pi}\lim_{m\rightarrow0}m^{2-2\Delta}\omega^{\Delta-1}\bigg(\frac{1}{\frac{m^2}{4\omega^2}+|z-z^{\prime}|^2}\bigg)^{2-\Delta}e^{\pm i\omega (2y\hat{p}^{\prime})\cdot X}\sim\lim_{m\rightarrow0}m^{2-2\Delta}\omega^{3-\Delta}=0
\end{split}
\end{align}
when $\omega=0$.
}
The proportional constant can be fixed by noting that
\begin{align}
\begin{split}
&(\Delta-1)\int d^2z \, m^{2\Delta-2}\bigg(\frac{1}{\frac{m^2}{4\omega^2}+|z-z^{\prime}|^2}\bigg)^{\Delta}=2^{2\Delta-1}\pi\omega^{2-2\Delta}\;,
\end{split}
\end{align}
which proves \eqref{eq:mLimit=delta_0}.

The two massless limits for the scalar wave functions can be summarized as
\ie
\phi_\Delta^\pm&=\lim_{m\to 0}\frac{m^\Delta(\Delta-1)}{2^\Delta\pi}\phi_{\Delta,m}^\pm\,,
\\
\widetilde\phi_{\Delta}^\pm&=\lim_{m\to 0}\frac{m^{2-\Delta}(1-\Delta)}{2^{2-\Delta}\pi}\phi_{\Delta,m}^\pm\,.
\fe

\section{Three-point celestial amplitudes of massless scalars}\label{sec:scalar_amplitudes}

In this section, we will compute the three-point celestial amplitudes involving three massless scalars. We will use the shadow basis \eqref{eq:phit} for outgoing particles and conformal primary basis \eqref{eq:Massless} for incoming particles. 
The computations of the one-to-two and the two-to-one amplitudes are in Section \ref{sec:2-1_amplitude} and \ref{sec:TwoOutgoing}, respectively.
For each amplitude, we give two computations that give the same results. The first computation is by directly performing the Mellin transform and shadow transform on the scattering amplitude in the plane wave basis, and the second computation is by taking the massless limit of the known celestial or shadow celestial amplitudes of one massive and two massless scalars.

\subsection{Two-to-one amplitude}\label{sec:2-1_amplitude}

In this subsection, we consider the $1^02^0\rightarrow3^0$ scattering. Here We use the superscript $0$ to indicate that the corresponding particle is a scalar.

\paragraph{Direct computation in momentum space} The corresponding celestial amplitude $\mathcal{A}_{1^02^0\rightarrow 3^0}^{\Delta_i}$ is given by 
\ie
\mathcal{A}_{1^02^0\rightarrow 3^0}^{\Delta_i}&=\int d^4 X\,\phi^+_{\Delta_1}(z_1;X)\phi^+_{\Delta_2}(z_2;X)\widetilde\phi^-_{\Delta_3}(z_3;X)
\\
&=\frac{(2\pi)^4\mathcal{N}_{\Delta_3,0}}{2^{1-\Delta_3}}\int_0^{\infty}d\omega_1d\omega_2\,\omega_1^{\Delta_1-1}\omega_2^{\Delta_2-1}\int\frac{d^3q^{\prime}_3}{q_3^{\prime0}}\frac{1}{(-\hat{q}_3\cdot q^{\prime}_3)^{\Delta_3}}\delta^{(4)}(q_1+q_2-q^{\prime}_3)\;,
\fe
where we have used \eqref{eq:Massless}, \eqref{eq:MasslessSpinl}, and \eqref{eq:MasslessScalarShadow} in the second equality.
By noting the identity 
\begin{align}
\int\frac{d^3q^{\prime}_3}{q^{\prime0}_3}=2\int d^4q^{\prime}_3\delta(-q^{\prime2}_3)\theta(q^{\prime0}_3)\;,
\end{align}
and using the delta-function $\delta^{(4)}(q_1+q_2-q^{\prime}_3)$, we can compute the integral over $d^4q^{\prime}_3$ leads to
\begin{align}
\mathcal{A}_{1^02^0\rightarrow 3^0}^{\Delta_i}=\frac{(2\pi)^4\mathcal{N}_{\Delta_3,0}}{2^{-\Delta_3}}\int_0^{\infty}d\omega_1d\omega_2\,\omega_1^{\Delta_1-1}\omega_2^{\Delta_2-1}\bigg(\frac{1}{-\hat{q}_3\cdot (q_1+q_2)}\bigg)^{\Delta_3}\delta(-2\omega_1\omega_2\hat{q}_1\cdot\hat{q}_2)\;.
\end{align}
The delta-function $\delta(-2\omega_1\omega_2\hat{q}_1\cdot\hat{q}_2)$ can be rewritten as
\begin{align}\label{eq:SupportDelta}
\delta(-2\omega_1\omega_2\hat{q}_1\cdot\hat{q}_2)=\frac{1}{2\omega_2(-\hat{q}_1\cdot\hat{q}_2)}\delta(\omega_1)+\frac{1}{2\omega_1(-\hat{q}_1\cdot\hat{q}_2)}\delta(\omega_2)+\frac{1}{2\omega_1\omega_2}\delta(-\hat{q}_1\cdot\hat{q}_2)\;.
\end{align}
We note that the first two terms in \eqref{eq:SupportDelta} correspond to the soft region and the last term corresponds to the colinear region. Using \eqref{eq:SupportDelta} and evaluating the integral over $\omega_1$ and $\omega_2$, we get 
\begin{align}\label{eq:ScalarOneOutgoing}
\mathcal{A}_{1^02^0\rightarrow 3^0}^{\Delta_i}=\mathcal{A}_{1^02^0\rightarrow 3^0}^{\text{soft},\Delta_i}+\mathcal{A}_{1^02^0\rightarrow 3^0}^{\text{colinear},\Delta_i}\;,
\end{align}
where $\mathcal{A}_{1^02^0\rightarrow 3^0}^{\text{soft},\Delta_i}$ and $\mathcal{A}_{1^02^0\rightarrow 3^0}^{\text{colinear},\Delta_i}$ are given by
\begin{align}\label{eq:ScalarOneOutgoingSoft}
\mathcal{A}_{1^02^0\rightarrow 3^0}^{\text{soft},\Delta_i}=(2\pi)^3\pi^2\mathcal{N}_{\Delta_3,0}(\frac{\delta_{\Delta_1,1}}{|z_{23}|^{2\Delta_3}}+\frac{\delta_{\Delta_2,1}}{|z_{13}|^{2\Delta_3}})\delta(\Delta_1+\Delta_2-\Delta_3-2)\frac{1}{|z_{12}|^2}\;,
\end{align}
and
\begin{align}\label{eq:ScalarOneOutgoingColinear}
\begin{split}
\mathcal{A}_{1^02^0\rightarrow 3^0}^{\text{colinear},\Delta_i}=&\frac{(2\pi)^5\mathcal{N}_{\Delta_3,0}\Gamma[\Delta_1-1]\Gamma[1-\Delta_1+\Delta_3]}{4\Gamma[\Delta_3]}\delta(\Delta_1+\Delta_2-\Delta_3-2)\frac{1}{|z_{13}|^{2\Delta_3}}\delta(|z_{12}|^2)\\
=&\frac{(2\pi)^3\pi^2\mathcal{N}_{\Delta_3,0}\Gamma[\Delta_1-1]\Gamma[\Delta_2-1]}{\Gamma[\Delta_3]}\delta(\Delta_1+\Delta_2-\Delta_3-2)\frac{1}{|z_{13}|^{2\Delta_3}}\delta(|z_{12}|^2)\;.
\end{split}
\end{align}
We mention here that the soft part $\mathcal{A}_{1^02^0\rightarrow 3^0}^{\text{soft},\Delta_i}$ (and the celestial amplitude $\mathcal{A}_{1^02^0\rightarrow 3^0}^{\Delta_i}$) converge only when $\Delta_1\geq1$ and $\Delta_2\geq1$.


\paragraph{Computation using massless limit}
\eqref{eq:ScalarOneOutgoing}, \eqref{eq:ScalarOneOutgoingSoft} and \eqref{eq:ScalarOneOutgoingColinear} can also be derived by taking the massless limit of the celestial amplitudes $\mathcal{A}^{\Delta_i,m}_{1^02^0\rightarrow 3^0}$ which involve two incoming massless scalars and one outgoing massive scalar with mass $m$. Specifically, according to \eqref{eq:MasslessScalarShadow}, \eqref{eqn:massive_scalar} and \eqref{eq:MasslessLimitGeneric}, we have
\begin{align}
\begin{split}
\mathcal{A}^{\Delta_i}_{1^02^0\rightarrow 3^0}=2^{\Delta_3-1}\lim_{m\rightarrow 0}\mathcal{A}^{\Delta_i,m}_{1^02^0\rightarrow3^0}\;.
\end{split}
\end{align}
We note that $\mathcal{A}^{\Delta_i,m}_{1^02^0\rightarrow3^0}$ is 
\ie\label{eq:MasslessMasslessMassive1}
\mathcal{A}_{1^02^0\rightarrow3^0}^{\Delta_i,m}&=\int d^4 X\, \phi^+_{\Delta_1}(z_1;X)\phi^+_{\Delta_2}(z_2;X)\Phi_m^{\Delta_3,-}(z_3;X)
\\
&=\frac{C^{\Delta_i}_{1^02^0\rightarrow3^0}(m)}{|z_{12}|^{\Delta_1+\Delta_2-\Delta_3}|z_{13}|^{\Delta_1-\Delta_2+\Delta_3}|z_{23}|^{\Delta_{2}+\Delta_3-\Delta_1}}\;,
\fe
where note that $\Phi_m^{\Delta,\pm}$ is related to $\phi^\pm_{\Delta,m}$ via \eqref{eqn:massive_scalar}.
Here the coefficient $C^{\Delta_i}_{1^02^0\rightarrow3^0}(m)$ is given by
\begin{align}\label{eq:Scalar3ptC1}
\begin{split}
C^{\Delta_i}_{1^02^0\rightarrow3^0}(m)=&2^{-\Delta_1-\Delta_2}(2\pi)^{4}\mathcal{N}_{\Delta_3,0}m^{\Delta_1+\Delta_2-\Delta_3-2}\frac{\Gamma[\frac{\Delta_{12}+\Delta_3}{2}]\Gamma[\frac{\Delta_3-\Delta_{12}}{2}]}{\Gamma[\Delta_3]}\\
=&2^{-\Delta_1-\Delta_2}(2\pi)^{4}\mathcal{N}_{\Delta_3,0}m^{a}\frac{\Gamma[-\frac{a}{2}+\Delta_1-1]\Gamma[-\frac{a}{2}+\Delta_2-1]}{\Gamma[\Delta_3]}\;,
\end{split}
\end{align}
where $a=\Delta_1+\Delta_2-\Delta_3-2$. We stress here that to get finite $\mathcal{A}_{1^02^0\rightarrow3^0}^{\Delta_i,m}$, 
\begin{align}\label{eq:ConvergenceCondition}
\begin{split}
\text{Re}(\frac{\Delta_{12}+\Delta_3}{2})=\text{Re}(-\frac{a}{2}+\Delta_1-1)\geq0\\
\text{Re}(\frac{-\Delta_{12}+\Delta_3}{2})=\text{Re}(-\frac{a}{2}+\Delta_2-1)\geq0
\end{split}
\end{align}
must hold. \footnote{These two conditions have to be satisfied such that the integral over $p_3$ is finite in the computation of $\mathcal{A}_{1^02^0\rightarrow3^0}^{\Delta_i,m}$.} Furthermore, to have well-defined massless limit, we need to assume $\text{Re}(a)\geq0$. Then taking the massless limit forces $-i\nu\equiv a=0$ with $\nu\in\mathbb{R}$. 

For generic $\Delta_i$s which satisfy \eqref{eq:ConvergenceCondition}, we can use
\begin{align}\label{eq:mLimit=delta}
\lim_{m\rightarrow0}m^{-i\nu}|z_{12}|^{i\nu-2}=2\lim_{m\rightarrow0}m^{-i\nu}\delta(|z_{12}|^2)\frac{1}{i\nu}=4\pi\delta(\nu)\delta(|z_{12}|^2)\;,    
\end{align}
where on the first equality we expanded $(|z_{12}|^2)^{\frac{i\nu}{2}-1}$ around $i\nu=0$ as \footnote{This follows from the distributional formula
\begin{align}
\begin{split}
x^a\theta(x)=\frac{(-1)^{n-1}\delta^{(n-1)}(x)}{(n-1)!}\frac{1}{a+n}+\cdots\,,
\end{split}
\end{align}
which is understood as follows. The LHS, when integrated against a test function, has poles on the complex $a$-plane at $a=-n$ for $n\in\bZ_{\ge0}$ with residues given by the residue of the RHS integrated against the same test function. More detailed discussions can be found in Appendix B of \cite{Caron-Huot:2022eqs} and references within.
}
\begin{align}
\begin{split}
(|z_{12}|^2)^{\frac{i\nu}{2}-1}&=2\delta(|z_{12}|^2)\frac{1}{i\nu}+\cdots\;.
\end{split}
\end{align}
Substituting \eqref{eq:mLimit=delta} into \eqref{eq:MasslessMasslessMassive1} then produces the colinear part $\mathcal{A}_{1^02^0\rightarrow 3^0}^{\text{colinear},\Delta_i}$ \eqref{eq:ScalarOneOutgoingColinear}.

On the other hand, for $\Delta_1=1$ or $\Delta_2=1$, we use the following formula 
\begin{align}\label{eq:mLimitGamma=delta}
\lim_{m\rightarrow0}m^{-i\nu}\Gamma[\frac{i\nu}{2}]=4\pi\delta(\nu)
\end{align}
to get
\begin{align}
\begin{split}
\lim_{m\rightarrow0}2^{\Delta_3-1}C^{\Delta_i}_{1^02^0\rightarrow3^0}(m)
=&(2\pi)^{3}\pi^2g\mathcal{N}_{\Delta_3,0}\delta(\nu)\;,
\end{split}
\end{align}
which agrees with the soft part $\mathcal{A}_{1^02^0\rightarrow 3^0}^{\text{soft},\Delta_i}$ \eqref{eq:ScalarOneOutgoingSoft}. The massless limit of $\mathcal{A}^{\Delta_i,m}_{1^02^0\rightarrow3^0}$ then can be obtained by adding the contribution from generic $\Delta_i$s and the contribution from $\Delta_1=1$ and $\Delta_2=1$ together. This reproduces $\mathcal{A}_{1^02^0\rightarrow 3^0}^{\Delta_i}$ \eqref{eq:ScalarOneOutgoing}.

\subsection{One-to-two amplitude}\label{sec:TwoOutgoing}

In this subsection, we consider the $3^0\rightarrow1^02^0$ scattering with three massless scalars.

\paragraph{Direct computaion in the momentum space} The corresponding celestial amplitude $\mathcal{A}_{3^0\rightarrow 1^02^0}^{\Delta_i}$ is given by 
\begin{align}
\begin{split}
\mathcal{A}^{\Delta_i}_{3^0\rightarrow 1^02^0}&=\int d^4 X\,\phi^+_{\Delta_1}(z_1;X)\widetilde\phi^-_{\Delta_2}(z_2;X)\widetilde\phi^-_{\Delta_3}(z_3;X)
\\
=&\frac{(2\pi)^4\mathcal{N}_{\Delta_1,0}\mathcal{N}_{\Delta_2,0}}{2^{2-\Delta_1-\Delta_2}}\int\frac{d^{3}q^{\prime}_1}{q^{\prime0}_1}\frac{d^{3}q^{\prime}_2}{q^{\prime0}_2}\int_0^{\infty}d\omega_3\,\omega_3^{\Delta_3-1} \frac{\delta^{(4)}(q^{\prime}_3-q^{\prime}_1-q^{\prime}_2)}{(-\hat{q}_1\cdot q^{\prime}_1)^{\Delta_1}(-\hat{q}_2\cdot q^{\prime}_2)^{\Delta_2}}\;,
\end{split}
\end{align}
where we have used \eqref{eq:Massless}, \eqref{eq:MasslessSpinl}, and \eqref{eq:MasslessScalarShadow} in the second equality.
Defining $p\equiv q^{\prime}_1+q^{\prime}_2\equiv M\hat{p}$ with $M\geq0$ and $\hat{p}^2=-1$ and changing integral variables then lead to \cite{Chang:2022jut}
\begin{align}
\begin{split}
\mathcal{A}^{\Delta_i}_{3^0\rightarrow 1^02^0}(z_i)=&\frac{\mathcal{N}_{\Delta_1,0}\mathcal{N}_{\Delta_2,0}}{2^{2-2\Delta_1-2\Delta_2}}\int_0^{+\infty}dM\int\frac{d^{3}\hat{p}}{\hat{p}^0}\int_0^{\infty}d\omega_3\,\omega_3^{\Delta_3-1}(2\pi)^4\delta^{(4)}(p-q^{\prime}_3)\\
&\qquad\times\int D^2\hat{q}^{\prime}_2\frac{M^{3-\Delta_1-\Delta_2}}{(-\hat{q}_2\cdot\hat{q}^{\prime}_2)^{\Delta_2}(-\hat{q}^{\prime}_2\cdot\hat{p})^{2-\Delta_1-\Delta_2}(-\hat{q}^{\prime}_2\cdot Y)^{\Delta_1}}\;,
\end{split}
\end{align}
where we defined $Y^{\mu}\equiv2(-\hat{q}_{1}\cdot\hat{p})\hat{p}^{\mu}-\hat{q}_{1}^{\mu}$. Using the following expansion
\begin{align}
\begin{split}
&\int D^2\hat{q}^{\prime}_2\frac{1}{(-\hat{q}_2\cdot\hat{q}^{\prime}_2)^{\Delta_2}(-\hat{q}^{\prime}_2\cdot\hat{p})^{2-\Delta_1-\Delta_2}(-\hat{q}^{\prime}_2\cdot Y)^{\Delta_1}}\\
&=\sum_{n=0}^{\infty}\frac{2\pi\Gamma[1-\Delta_1]\Gamma[1-\Delta_2](\Delta_1)_n(\Delta_2)_n}{2^{\Delta_1+\Delta_2}\Gamma[2-\Delta_1-\Delta_2]\Gamma[n+1]\Gamma[n+1]}\frac{(\frac{1}{2}\hat{q}_{1}\cdot\hat{q}_2)^n}{(\hat{q}_1\cdot\hat{p})^{\Delta_1+n}(-\hat{q}_2\cdot\hat{p})^{\Delta_2+n}}\;,
\end{split}
\end{align}
we get
\begin{align}
\begin{split}
\widetilde{\mathcal{A}}^{\Delta_i}_{3^0\rightarrow 1^02^0}=&\mathcal{N}_{\Delta_1,0}\mathcal{N}_{\Delta_2,0}\frac{(2\pi)^5\Gamma[1-\Delta_1]\Gamma[1-\Delta_2]}{2^{2-\Delta_1-\Delta_2}\Gamma[2-\Delta_1-\Delta_2]}\sum_{n=0}^{\infty}\frac{(\Delta_1)_n(\Delta_2)_n}{(n!)^2}\int_0^{+\infty}dMM^{3-\Delta_1-\Delta_2}\\
&\times (\frac{1}{2}\hat{q}_{1}\cdot\hat{q}_2)^n\int\frac{d^{3}\hat{p}}{\hat{p}^0}\int_0^{\infty}d\omega_3\,\omega_3^{\Delta_3-1}\frac{\delta^{(4)}(p-q^{\prime}_3)}{(-\hat{q}_1\cdot \hat{p}_1)^{\Delta_1+n}(-\hat{q}_2\cdot\hat{p}_2)^{\Delta_2+n}}\;.
\end{split}
\end{align}
Defining $\lambda$, $y$, $z$, and $\bar{z}$ through
\begin{align}
M\hat{p}=\lambda(1+y^2+z\bar{z},z+\bar{z},-i(z-\bar{z}),1-y^2-z\bar{z})
\end{align}
leading to
\begin{align}
\begin{split}
\mathcal{A}^{\Delta_i}_{3^0\rightarrow 1^02^0}=&\mathcal{N}_{\Delta_1,0}\mathcal{N}_{\Delta_2,0}\frac{2(2\pi)^5\Gamma[1-\Delta_1]\Gamma[1-\Delta_2]}{\Gamma[2-\Delta_1-\Delta_2]}\sum_{n=0}^{\infty}\frac{(\Delta_1)_n(\Delta_2)_n}{(n!)^2}\int_0^{+\infty}d\lambda\;\lambda^{3-\Delta_1-\Delta_2}\\
&\times (\frac{1}{2}\hat{q}_{1}\cdot\hat{q}_2)^n\int_0^{\infty}dy\;y^{2n+1}\int d^2z\bigg(\frac{1}{|z_1-z|^2-y^2}\bigg)^{\Delta_1+n}\bigg(\frac{1}{|z_2-z|^2-y^2}\bigg)^{\Delta_2+n}\\
&\times\int_0^{\infty}d\omega_3\,\omega_3^{\Delta_3-1}\delta^{(4)}(2y\lambda\hat{p}-q^{\prime}_3)\;.
\end{split}
\end{align}
Rewriting the delta-function as
\begin{align}
\delta^{(4)}(2y\lambda\hat{p}-p^{\prime}_3)=\frac{1}{8y\lambda\omega_3^2}\delta(y)\delta(\lambda-\omega_3)\delta^{(2)}(z_3-z) 
\end{align}
and evaluating the remaining integrals leads to
\begin{align}\label{eq:ScalarTwoOutgoing}
\begin{split}
\mathcal{A}^{\Delta_i}_{3^0\rightarrow 1^02^0}=&\mathcal{N}_{\Delta_1,0}\mathcal{N}_{\Delta_2,0}\frac{(2\pi)^4\pi^2\Gamma[1-\Delta_1]\Gamma[1-\Delta_2]}{\Gamma[2-\Delta_1-\Delta_2]}\delta(\Delta_1+\Delta_2-\Delta_3)\frac{1}{|z_{13}|^{2\Delta_1}|z_{23}|^{2\Delta_2}}\;.
\end{split}
\end{align}
%


\paragraph{Computation using massless limit}
\eqref{eq:ScalarTwoOutgoing} can also be derived by taking the massless limit of the celestial amplitudes $\mathcal{A}^{\Delta_i,m}_{1\rightarrow2}$ which involve two outgoing massless particles and one incoming massive particle with mass $m$. Specifically, according to \eqref{eq:MasslessLimitScalar1}, we have
\begin{align}\label{eq:A1t2=Am}
\begin{split}
\mathcal{A}^{\Delta_i}_{3^0\rightarrow 1^02^0}=-2^{1-\Delta_3}\lim_{m\rightarrow 0}m^{2\Delta_3-2}\mathcal{A}^{\Delta_i,m}_{3^0\rightarrow1^02^0}\;.
\end{split}
\end{align}
$\mathcal{A}^{\Delta_i,m}_{3^0\rightarrow1^02^0}$ has been computed in \cite{Chang:2022jut}, given by
\ie\label{eq:MasslessMasslessMassiveTwoOutgoing}
\mathcal{A}_{3^0\rightarrow1^02^0}^{\Delta_i,m}
&=\int d^4 X\, \Phi_m^{\Delta_3,+}(z_3;X)\phi^-_{\Delta_1}(z_1;X)\phi^-_{\Delta_2}(z_2;X)
\\
&=\frac{C^{\Delta_i}_{3^0\rightarrow1^02^0}(m)}{|z_{12}|^{\Delta_1+\Delta_2-\Delta_3}|z_{13}|^{\Delta_1-\Delta_2+\Delta_3}|z_{23}|^{\Delta_{2}+\Delta_3-\Delta_1}}\;,
\fe
where the third particle is massive (note that $\Phi_m^{\Delta,\pm}$ is related to $\phi^\pm_{\Delta,m}$ via \eqref{eqn:massive_scalar}) and the coefficient $C^{\Delta_i}_{3^0\rightarrow1^02^0}(m)$ is given by
\begin{align}\label{eq:Scalar3ptC}
\begin{split}
C^{\Delta_i}_{3^0\rightarrow1^02^0}(m)=&(2\pi)^{4}\pi^2 \mathcal{N}_{\Delta_1,0}\mathcal{N}_{\Delta_2,0}\mathcal{N}_{\Delta_3,0}m^{2-\Delta_1-\Delta_2-\Delta_3}\\
&\times\frac{\Gamma[1-\Delta_1]\Gamma[1-\Delta_2]\Gamma[\frac{\Delta_1+\Delta_{23}}{2}]\Gamma[\frac{\Delta_{12}+\Delta_3}{2}]\Gamma[\frac{\Delta_3-\Delta_{12}}{2}]\Gamma[\frac{\sum_{i=1}^3\Delta_i-2}{2}]}{2^{2-\Delta_1-\Delta_2}\Gamma[\Delta_1]\Gamma[\Delta_2]\Gamma[\Delta_3]\Gamma[\frac{2-\Delta_1-\Delta_{23}}{2}]\Gamma[\frac{4-\sum_{i=1}^3\Delta_i}{2}]}\;.
\end{split}
\end{align}
After assuming $\text{Re}(\Delta_3-\Delta_1-\Delta_2)\geq0$ and using \eqref{eq:mLimitGamma=delta}, we find that
\ie
-2^{1-\Delta_3}\lim_{m\rightarrow 0}\bigg(m^{2\Delta_3-2}C^{\Delta_i}_{3^0\rightarrow1^02^0}(m)\bigg)=&\mathcal{N}_{\Delta_1,0}\mathcal{N}_{\Delta_2,0}\frac{g(2\pi)^4\pi^2\Gamma[1-\Delta_{1}]\Gamma[1-\Delta_2]}{\Gamma[2-\Delta_1-\Delta_2]}
    \\
    &\times \delta(\Delta_1+\Delta_2-\Delta_3)\;,
\fe
which agrees with \eqref{eq:ScalarTwoOutgoing}.

\section{Three-point celestial amplitudes with a massless spinning particle}\label{sec:l-0-0_amplitudes}
In this section, we consider the following three-point scattering amplitudes
\begin{align}\label{eq:MPhotonScalarScalar}
\mathcal{M}^{\mu}=(2\pi^4)\delta^{(4)}(q_1-q_2-q_3)(q^{\mu}_1+q^{\mu}_3)
\end{align}
for scalar-photon-scalar and
\begin{align}
\begin{split}
\mathcal{M}_{\mu\nu}=(2\pi^4)\frac{1}{2}\delta^{(4)}(q_1-q_2-q_3)(q_{1\mu}q_{3\nu}+q_{1\nu}q_{3\mu}-\eta_{\mu\nu}q_1\cdot q_3)
\end{split}
\end{align}
for scalar-graviton-scalar. 

\subsection{Scalar-photon-scalar amplitude}\label{sec:PhotonScalarScalar}

Let us first compute the celestial amplitude $\mathcal{A}^{\Delta_i}_{1^0\rightarrow2^-3^0}$ of one incoming massless scalar, one outgoing photon, and one outgoing massless scalar. \footnote{The helicity $+$ or $-$ in this paper indicates the helicity in four-dimensional Minkowski space. For incoming particles, the helicity in two-dimensional CCFT concides with the helicity in four-dimensional Minkowski space. On the other hand,  for outgoing particles, the helicity in two-dimensional CCFT get flipped comparing with the helicity in four-dimensional Minkowski space. This is because that the shadow transformation flips the helicity.} 
Using \eqref{eq:Massless}, \eqref{eq:SpinOneMellin}, \eqref{eq:phit} and \eqref{eq:MPhotonScalarScalar}, we find that the celestial amplitude $\mathcal{A}^{\Delta_i}_{1^0\rightarrow2^-3^0}$ is \footnote{As we will see later, only the soft mode with $\Delta_2=1$ has contribution. Since $A^{\Delta_2,\pm}_{\mu z}=\widetilde{A}^{\Delta_2,\pm}_{\mu z}$ when $\Delta_2=1$, we can use the conformal primary basis $A^{\Delta_2,\pm}_{\mu z}$ although photon is outgoing.}
\begin{align}
\begin{split}
\mathcal{A}^{\Delta_i}_{1^0\rightarrow2^-3^0}=&\frac{(2\pi^4)\mathcal{N}_{\Delta_3,0}}{2^{1-\Delta_3}}\frac{(\Delta_2-1)}{\Delta_2N^-_{\Delta_2}}\partial_{z_2}\hat{q}_2^{\mu}\int_{0}^{\infty}d\omega_1d\omega_2\,\omega_1^{\Delta_1-1}\omega_2^{\Delta_2-1}\\
&\qquad\times\int\frac{d^3q^{\prime}_3}{q^{\prime0}_3}\frac{1}{(\hat{q}_3\cdot q_3^{\prime})^{\Delta_3}}\delta^{(4)}(q_1-q_2-q^{\prime}_3)(q^{\mu}_1+q^{\prime\mu}_3)\;,
\end{split}
\end{align}
where we used the momentum conservation to rewrite $q^{\mu}_1+q^{\prime\mu}_3$ as $2q^{\mu}_1-q^{\mu}_2$ and dropped $q_2^{\mu}$ since $\partial_{z_2}\hat{q}_2\cdot q_2=0$. As we have seen in \eqref{eq:SupportDelta}, the supports of momentum conservation contain the soft regions and the colinear region:
\begin{align}
\begin{split}
&\omega_3=0,\qquad q_1=q_2\;,\\
&\omega_2=0,\qquad q_1=q^{\prime}_3\;,\\
&\omega_1+\omega_2=\omega_3,\qquad\hat{q}_1=\hat{q}_2=\hat{q}^{\prime}_3\;.
\end{split}
\end{align}
However, since the colinear region demands $\hat{q}_1=\hat{q}_2=\hat{q}^{\prime}_3$ and the conformal primary basis is transverse with respect to the momentum, the colinear region has no contribution to the celestial amplitude. Moreover, we note that the soft region $\omega_3=0$ implies $q^{\mu}_1=q^{\mu}_2$, which leads to vanishing celestial amplitude. Thus, only the soft region $\omega_2=0$ has non-vanishing contribution to the celestial amplitude $\mathcal{A}^{\Delta_i}_{1^0\rightarrow2^+3^0}$ and we can write the delta-function as
\begin{align}
\delta^{(4)}(q_1-q_2-q^{\prime}_3)=\frac{q^{\prime0}_3}{\omega_3\hat{q}_2\cdot \hat{q}_3^{\prime}}\delta(\omega_2)\delta^{(3)}(q_1-q^{\prime}_3)\;.
\end{align}
This leads to
\begin{align}
\begin{split}
\mathcal{A}^{\Delta_i}_{1^0\rightarrow2^-3^0}=&2(2\pi^4)\mathcal{N}_{\Delta_3,0}\frac{(\Delta_2-1)}{ \Delta_2N^-_{\Delta_2}}\partial_{z_2}\hat{q}_2^{\mu}\int_{0}^{\infty}d\omega_1d\omega_2\,\omega_1^{\Delta_1-2}\omega_2^{\Delta_2-1}\delta(\omega_2)\\
&\qquad\times\int d^3q^{\prime}_3\frac{1}{(\hat{q}_3\cdot q_3^{\prime})^{\Delta_3}}\frac{1}{\hat{q}_2\cdot \hat{q}_3^{\prime}}\delta^{(3)}(q_1-q^{\prime}_3)q^{\mu}_1\;.
\end{split}
\end{align}
To have convergent $\omega_2$-integral, we have to assume $\text{Re}(\Delta_2)\geq1$. Then the $\omega_2$-integral forces $\Delta_2=1$ which vanishes due to the prefactor $\Delta_2-1$. Thus we conclude that $\mathcal{A}^{\Delta_i}_{1^0\rightarrow2^-3^0}$ vanishes. The vanishing of $\mathcal{A}^{\Delta_i}_{1^0\rightarrow2^-3^0}$ is due to the soft mode of photon with $\Delta_2=1$. 

As we discussed in Section \ref{sec:ConformalBasis}, instead of using \eqref{eq:SpinOne}, we should use \eqref{eq:Alog} as the spin-one conformal primary basis for the soft mode with $\Delta_2=1$. In other words, the celestial amplitude $\mathcal{A}^{\Delta_i}_{1^0\rightarrow2^-_{\text{soft}}3^0}$ should be computed from
\begin{align}
\begin{split}
\mathcal{A}^{\Delta_i}_{1^0\rightarrow2^-_{\text{soft}}3^0}=&\frac{(2\pi^4)\mathcal{N}_{\Delta_3,0}}{2^{1-\Delta_3}}\int_{0}^{\infty}d\omega_1\,\omega_1^{\Delta_1-1}\int\frac{d^3q^{\prime}_2}{q^{\prime0}_2}A^{\text{log},-}_{\mu z}(\hat{q}_2;q_2^{\prime})\\
&\qquad\times\int\frac{d^3q^{\prime}_3}{q^{\prime0}_3}\frac{1}{(\hat{q}_3\cdot q_3^{\prime})^{\Delta_3}}\delta^{(4)}(q_1-q^{\prime}_2-q^{\prime}_3)(q^{\mu}_1+q^\mu_3)\;.
\end{split}
\end{align}
Here $A^{\text{log},-}_{\mu z}(\hat{q}_2;q_2^{\prime})$ is $A^{\text{log},-}_{\mu z}(\hat{q}_2;X)$ \eqref{eq:Alog} in momentum space, defined similar to \eqref{eq:SpinOneFourier}.
Re-writing the delta-function $\delta^{(4)}(q_1-q^{\prime}_2-q^{\prime}_3)$ as
\begin{align}
\delta^{(4)}(q_1-q^{\prime}_2-q^{\prime}_3)=\frac{1}{(2\pi)^4}\int d^4Xe^{i(q_1-q^{\prime}_2-q^{\prime}_3)\cdot X}    
\end{align}
leads to
\begin{align}
\begin{split}
\mathcal{A}^{\Delta_i}_{1^0\rightarrow2^-_{\text{soft}}3^0}=&\frac{-i\mathcal{N}_{\Delta_3,0}}{2^{1-\Delta_3}}\int d^4X\int\frac{d^3q^{\prime}_2}{q^{\prime0}_2}A^{\text{log},-}_{\mu z}(\hat{q}_2;q_2^{\prime})e^{-iq^{\prime}_2\cdot X}
\\
&~\times\bigg(\frac{\partial }{\partial X_{1,\mu}}-\frac{\partial }{\partial X_{3,\mu}}\bigg)\int_{0}^{\infty}d\omega_1\,\omega_1^{\Delta_1-1}e^{iq_1\cdot X_1}\int\frac{d^3q^{\prime}_3}{q^{\prime0}_3}\frac{1}{(\hat{q}_3\cdot q_3^{\prime})^{\Delta_3}}e^{-iq^{\prime}_3\cdot X_3}\bigg|_{X_1,X_3\to X}\;.
\end{split}
\end{align}
Computing the integral over $d\omega_1$, $d^3q^{\prime}_2$ and $d^3q^{\prime}_3$, we get the integral representation of $\mathcal{A}^{\Delta_i}_{1^0\rightarrow2^-_{\text{soft}}3^0}$ in the position space: \footnote{In the rest of paper, we do not write $(\hat{q};X)$ explicitly for conformal primary basis in position space. For example, $\phi_{\Delta}^{\pm}$ is used to denote $\phi_{\Delta}^{\pm}(\hat{q};X)$.}
\begin{align}
\begin{split}
\mathcal{A}^{\Delta_i}_{1^0\rightarrow2^-_{\text{soft}}3^0}=&-i\int d^4XA^{\text{log},-}_{2,\mu z}\left[\partial^{\mu}\phi_{1,\Delta_1}^{+}\widetilde{\phi}^{-}_{3,\Delta_3}- \phi_{1,\Delta_1}^{+}\partial^{\mu}\widetilde{\phi}^{-}_{3,\Delta_3}\right]\\
=&-i\int d^4X[V^{\text{log},-}_{2,\mu z}+\partial_{\mu}\alpha^{\text{log},-}_{2,z}]\left[\partial^{\mu}\phi_{1,\Delta_1}^{+}\widetilde{\phi}^{-}_{3,\Delta_3}- \phi_{1,\Delta_1}^{+}\partial^{\mu}\widetilde{\phi}^{-}_{3,\Delta_3}\right]\;,
\end{split}
\end{align}
where we used \eqref{eq:Alog}, and use the abbreviations $\partial_{\mu}\equiv\frac{\partial}{\partial X^{\mu}}$, $A^{\text{log},-}_{2,\mu z}\equiv A^{\text{log},-}_{\mu z}(\hat q_2;X)$ and similarly for other fields. Using the fact that $\partial^2\phi=\partial^2\widetilde{\phi}=0$, we have
\begin{align}
\begin{split}
&\int d^4X\partial_{\mu}\alpha^{\text{log},-}_{2,z}\left[\partial^{\mu}\phi_{1,\Delta_1}^{+}\widetilde{\phi}^{-}_{3,\Delta_3}- \phi_{1,\Delta_1}^{+}\partial^{\mu}\widetilde{\phi}^{-}_{3,\Delta_3}\right]
\\
&=\int d^4X\alpha^{\text{log},-}_{2,z}\left[-\partial^{\mu}\phi_{1,\Delta_1}^{+}\partial_{\mu}\widetilde{\phi}^{-}_{3,\Delta_3}+\partial^{\mu}\phi_{1,\Delta_1}^{+}\partial_{\mu}\widetilde{\phi}^{-}_{3,\Delta_3}\right]\\
&=0\;.
\end{split}
\end{align}
This leads to
\begin{align}\label{eq:PhotonPositionSpcae}
\begin{split}
\mathcal{A}^{\Delta_i}_{1^0\rightarrow2^-_{\text{soft}}3^0}=&-i\int d^4XV^{\text{log},-}_{2,\mu z}\left[\partial^{\mu}\phi_{1,\Delta_1}^{+}\widetilde{\phi}^{-}_{3,\Delta_3}- \phi_{1,\Delta_1}^{+}\partial^{\mu}\widetilde{\phi}^{-}_{3,\Delta_3}\right]\;.
\end{split}
\end{align}
Using \eqref{eq:Massless}, \eqref{eq:phit=phi} and \eqref{eq:Alog}, $\mathcal{A}^{\Delta_i}_{1\rightarrow2,z}$ can be written as
\begin{align}
\begin{split}
\mathcal{A}^{\Delta_i}_{1^0\rightarrow2^-_{\text{soft}}3^0}=&-i(2-\Delta_1-\Delta_3)N^{+}_{\Delta_1}N^{-}_{2-\Delta_3}\int d^4X\frac{1}{(-\hat{q}_1\cdot X)^{\Delta_1}}\frac{1}{X^2}\frac{\partial_{z_2}\hat{q}_2\cdot X}{(-\hat{q}_2\cdot X)}\frac{(-X^2)^{\Delta_3-1}}{(-\hat{q}_3\cdot X)^{\Delta_3}}\\
=&i(2-\Delta_1-\Delta_3)N^{+}_{\Delta_1}N^{-}_{2-\Delta_3}\\
&\qquad\times\lim_{\Delta_2^{\prime}\rightarrow0}\frac{1}{\Delta_2^{\prime}}\partial_{z_2}\int d^4X\frac{1}{(-\hat{q}_1\cdot X)^{\Delta_1}}\frac{(-X^2)^{\Delta_2^{\prime}-1}}{(-\hat{q}_2\cdot X)^{\Delta_2^{\prime}}}\frac{(-X^2)^{\Delta_3-1}}{(-\hat{q}_3\cdot X)^{\Delta_3}}\;.
\end{split}
\end{align}
We recognize that the $X$-integral is just the celestial amplitude $\mathcal{A}_{1^0\rightarrow2^03^0}^{\Delta_3\Delta_2^{\prime}\Delta_1}$ which can be obtained from \eqref{eq:ScalarTwoOutgoing} by switching $1$ and $3$, \textit{i.e.,}
\begin{align}\label{eq:eq1}
\begin{split}
\mathcal{A}^{\Delta_i}_{1^0\rightarrow2^-_{\text{soft}}3^0}=&2i(2-\Delta_1-\Delta_3)\lim_{\Delta_2^{\prime}\rightarrow0}\frac{1}{{N}^{-}_{2-\Delta_2'}\Delta_2^{\prime}}\partial_{z_2}\int d^4X\phi^{+}_{1,\Delta_1}\widetilde{\phi}_{2,\Delta_2^{\prime}}^{-}\widetilde{\phi}^{-}_{3,\Delta_3}\\
=&2i(2-\Delta_1-\Delta_3)\lim_{\Delta_2^{\prime}\rightarrow0}\frac{1}{{N}^{-}_{2-\Delta_2'}\Delta_2^{\prime}}\partial_{z_2}\mathcal{A}^{\Delta_3\Delta_2^{\prime}\Delta_1}_{1^0\rightarrow2^03^0}\;.
\end{split}
\end{align}

Now we reach a subtlety. Substituting the expression \eqref{eq:ScalarTwoOutgoing} into the above equality, we will get something that does not transform correctly under the conformal symmetry. This is because \eqref{eq:ScalarTwoOutgoing} contains the delta-function $\delta(\Delta_1+\Delta_2-\Delta_3)$ that is singular and needs to be regularized while preserving the conformal symmetry. One resolution is to turn on a small mass $m$ for the incoming particle, which smooths out the delta-function while keeping the conformal symmetry manifestly. We would take the massless limit $m\rightarrow0$ at the end of the computation. In other words, we have
\begin{align}
\begin{split}
\mathcal{A}^{\Delta_i}_{1^0\rightarrow2^-_{\text{soft}}3^0}=&2i(2-\Delta_1-\Delta_3)\lim_{m\rightarrow0}\bigg(-2^{1-\Delta_1}m^{2\Delta_1-2}\lim_{\Delta_2^{\prime}\rightarrow0}\frac{1}{{N}^{-}_{2-\Delta_2'}\Delta_2^{\prime}}\partial_{z_2}\mathcal{A}^{\Delta_3\Delta_2^{\prime}\Delta_1,m}_{1^0\rightarrow2^03^0}\bigg)\;,
\end{split}
\end{align}
where we used \eqref{eq:A1t2=Am} and $\mathcal{A}^{\Delta_3\Delta_2^{\prime}\Delta_1,m}_{1^0\rightarrow2^03^0}$ is obtained from \eqref{eq:MasslessMasslessMassiveTwoOutgoing} by switching $1$ and $3$. Substituting \eqref{eq:MasslessMasslessMassiveTwoOutgoing} and \eqref{eq:Scalar3ptC} into the above equality and using \eqref{eq:mLimitGamma=delta} leads to
\begin{align}\label{eq:PhotonScalarScalar}
\begin{split}
\mathcal{A}^{\Delta_i}_{1^0\rightarrow2^-_{\text{soft}}3^0}
=&(2\pi)^4i(\Delta_1-1)\delta(\Delta_1-\Delta_3)(\frac{1}{z_{12}}+\frac{1}{z_{23}})\frac{1}{|z_{13}|^{2\Delta_1}}\;,
\end{split}
\end{align}
which has the standard form of scalar-current-scalar three-point functions in 2d CFT. 

We stress here that to get \eqref{eq:PhotonScalarScalar}, we introduced a mass regulator. In Appendix \ref{sec:ShadowAmplitude}, we give another way to compute \eqref{eq:PhotonScalarScalar} by performing the shadow transformation on the shadow celestial amplitude $\widetilde{\mathcal{A}}^{\Delta_i}_{1^0\rightarrow2^-_{\text{soft}}3^0}$.

\subsection{Scalar-graviton-scalar amplitude}

Let us compute the celestial amplitude $\mathcal{A}^{\Delta_i}_{1^0\rightarrow2^{--}3^0}$ of one incoming massless scalar, one outgoing graviton, and one outgoing massless scalar. Like the case of scalar-photon-scalar, only the soft region $\omega_2=0$ contributes to the celestial amplitude $\mathcal{A}^{\Delta_i}_{1^0\rightarrow2^{--}3^0}$. Again the delta-function $\delta(\omega_2)$ forces $\Delta_2=1$ which implies that we should use $h^{\text{log},-}_{2,\mu\nu zz}$ in \eqref{eq:hlog} to compute $\mathcal{A}^{\Delta_i}_{1^0\rightarrow2^{--}_{\text{soft}}3^0}$. Following the computation in Section \ref{sec:PhotonScalarScalar},
we get
\begin{align}
\begin{split}
\mathcal{A}^{\Delta_i}_{1^0\rightarrow2^{--}_{\text{soft}}3^0}=&\frac{1}{2}\int d^4X\;h^{\text{log},-}_{2,\mu\nu zz}\left[\partial^{\mu}\phi_{1,\Delta_1}^{+}\partial^{\nu}\widetilde{\phi}^{-}_{3,\Delta_3}+\partial^{\nu}\phi_{1,\Delta_1}^{+}\partial^{\mu}\widetilde{\phi}^{-}_{3,\Delta_3}-g^{\mu\nu}\partial\phi_{1,\Delta_1}^{+}\cdot\partial\widetilde{\phi}^{-}_{3,\Delta_3}\right]\\
=&\frac{1}{2}\int d^4X\;[W^{\text{log},-}_{2,\mu\nu zz}+\partial_{(\mu}\alpha^{\text{log},-}_{2,\nu)zz}]\\
&\qquad\times\left[\partial^{\mu}\phi_{1,\Delta_1}^{+}\partial^{\nu}\widetilde{\phi}^{-}_{3,\Delta_3}+\partial^{\nu}\phi_{1,\Delta_1}^{+}\partial^{\mu}\widetilde{\phi}^{-}_{3,\Delta_3}-g^{\mu\nu}\partial\phi_{1,\Delta_1}^{+}\cdot\partial\widetilde{\phi}^{-}_{3,\Delta_3}\right]\;.
\end{split}
\end{align}
Since $\phi_{1,\Delta_1}^{+}$ and $\widetilde{\phi}^{-}_{3,\Delta_3}$ satisfy the massless Klein-Gordon equation, the term including $\partial_{(\mu}\alpha^{\text{log},-}_{2,\nu)zz}$ vnishes after integrating by parts, leading to
\begin{align}
\begin{split}
\mathcal{A}^{\Delta_i}_{1^0\rightarrow2^{--}_{\text{soft}}3^0}=&\frac{1}{2}\int d^4X\;W^{\text{log},-}_{2,\mu\nu zz}\left[\partial^{\mu}\phi_{1,\Delta_1}^{+}\partial^{\nu}\widetilde{\phi}^{-}_{3,\Delta_3}+\partial^{\nu}\phi_{1,\Delta_1}^{+}\partial^{\mu}\widetilde{\phi}^{-}_{3,\Delta_3}-g^{\mu\nu}\partial\phi_{1,\Delta_1}^{+}\cdot\partial\widetilde{\phi}^{-}_{3,\Delta_3}\right]\;.
\end{split}
\end{align}
Using \eqref{eq:WLog}, \eqref{eq:Massless} and \eqref{eq:phit=phi}, $\mathcal{A}^{\Delta_i}_{1^0\rightarrow2^{--}_{\text{soft}}3^0}$ bceomes that
\begin{align}
\begin{split}
\mathcal{A}^{\Delta_i}_{1^0\rightarrow2^{--}_{\text{soft}}3^0}=&\Delta_1(2-\Delta_3) N_{\Delta_1}^{+}N_{2-\Delta_3}^-\int d^4X\frac{(\partial_{z_2}\hat q_2\cdot X)^2}{(-\hat q_2\cdot X)}\frac{1}{(-\hat{q}_1\cdot X)^{\Delta_1}}\frac{(-X^2)^{\Delta_3-3}}{(-\hat{q}_3\cdot X)^{\Delta_3}}\\
=&\lim_{\Delta_2^{\prime}\rightarrow-1}\frac{\Delta_1(2-\Delta_3) }{N^{-}_{2-\Delta_2^{\prime}}\Delta_2^{\prime}(\Delta_2^{\prime}+1)}\partial_{z_2}\partial_{z_2}\mathcal{A}^{\Delta_3\Delta_2^{\prime}\Delta_1}_{1^0\rightarrow2^03^0}\;.
\end{split}
\end{align}
Following the computations in Section \ref{sec:PhotonScalarScalar} with the mass regulator, we finally get that
\begin{align}\label{eq:ScalarGravitonScalar}
\begin{split}
\mathcal{A}^{\Delta_i}_{1^0\rightarrow2^{--}_{\text{soft}}3^0}=&4i\pi^4\Delta_1\delta(\Delta_3-\Delta_1-1)\frac{\bar z_{12}z_{13}^2}{z_{12} z_{23}^2}\frac{1}{|z_{13}|^{2\Delta_1+2}}\;,
\end{split}
\end{align}
which is a three-point function of primary operators with conformal weights $(\frac{\Delta_1}{2},\frac{\Delta_1}{2})$, $(\frac{3}{2},-\frac{1}{2})$, $(\frac{\Delta_3}{2},\frac{\Delta_3}{2})$. Following \cite{Donnay:2018neh}, we can further define an operator $P_z$ with conformal weight $(\frac{3}{2},\frac{1}{2})$, which was referred to as the supertranslation current \cite{Strominger:2013jfa,Barnich:2013axa}. By acting $\partial_{\bar{z}_2}$ on \eqref{eq:ScalarGravitonScalar}, we obtain the three-point function $\langle\mathcal{O}_{\Delta_1}^{+}(z_1)P_{z}(z_2)\mathcal{O}_{\Delta_3}^{-}(z_3)\rangle$:
\begin{align}\label{eq:ScalarPScalar}
\begin{split}
\langle\mathcal{O}_{\Delta_1}^{+}(z_1)P_{z}(z_2)\mathcal{O}_{\Delta_3}^{-}(z_3)\rangle=&-4i\pi^4\Delta_1\delta(\Delta_3-\Delta_1-1)\frac{z_{13}^2}{z_{12} z_{23}^2}\frac{1}{|z_{13}|^{2\Delta_1+2}}\;.
\end{split}
\end{align}
Thus the OPE between $P_z$ and $\mathcal{O}_{\Delta_1}^{+}(z_1)$ takes the form as
\begin{align}\label{eq:OPESupertranslation}
P_z(z_2)\mathcal{O}_{\Delta_1}^{+}(z_1)\sim \frac{-4i\pi^4(\Delta_3-1)\delta(\Delta_3-\Delta_1-1)}{C^{\Delta_1+1,\Delta_3}}\frac{1}{z_{12}}\mathcal{O}_{\Delta_3}^{-}(z_2)\;,
\end{align}
where $C^{\Delta_1+1,\Delta_3}$ is the coefficient of the two-point celestial amplitude of massless scalars, which contains a delta function $\delta(\Delta_3-\Delta_1-1)$ that cancels the delta function in the numerator. \eqref{eq:OPESupertranslation} agrees with the result in \cite{Donnay:2018neh}.

Finally, by a computation similar to the one in Section \ref{sec:scalar_amplitudes}, one may alternatively obtain the scalar-photon-scalar and the scalar-graviton-scalar celestial amplitudes by taking the massless limit of the massive spinning particle amplitudes in \cite{Chang:2021wvv}. We leave this to future work.

\section{Three-point celestial gluon amplitudes}\label{sec:l-l-l_amplitudes}

In this section, we first compute the celestial amplitude $\mathcal{A}^{\Delta_i}_{1^a\rightarrow 2^b3^c}$ of three gluons with helicities $(a,b,c)$ in Minkowski space. Here, we omit the color indices. After that, we compare our results with the existing three-gluon celestial amplitude in the Klein space \cite{Pasterski:2017ylz}. 

\subsection{Computation of $\mathcal{A}^{\Delta_i}_{1^a\rightarrow 2^b3^c}$}

We start with the corresponding scattering amplitude $\mathcal{M}^{\mu\nu\rho}$ which takes the form as \footnote{Here we omit the structure constant $f^{abc}$.}
\begin{align}\label{eq:MGluonGluonGluon}
\mathcal{M}^{\mu\nu\rho}=(2\pi^4)\delta^{(4)}(q_1-q_2-q^{\prime}_3)(g^{\mu\nu}(q_1+q_2)^{\rho}+g^{\nu\rho}(-q_2+q^{\prime}_3)^{\mu}+g^{\rho\mu}(-q^{\prime}_3-q_1)^{\nu})\;.
\end{align}
Since $z_i$ and $\bar{z}_i$ are not independent in Minkowski space, the colinear region dose not contribute to the celestial amplitude $\mathcal{A}^{\Delta_i}_{1^a\rightarrow 2^b3^c}$. However, we still have contribution from the soft region. Indeed, we can write the delta-function for momentum conservation as
\begin{align}\label{eq:delta=omega2+omega3}
\delta^{(4)}(q_1-q_2-q^{\prime}_3)=\frac{q^{\prime0}_3}{\omega_3\hat{q}_2\cdot \hat{q}_3^{\prime}}\delta(\omega_2)\delta^{(3)}(q_1-q^{\prime}_3)+\frac{q^{0}_2}{\omega_2\hat{q}_2\cdot \hat{q}_3^{\prime}}\delta(\omega_3)\delta^{(3)}(q_1-q_2)\;.
\end{align}
Following the steps in Section \ref{sec:PhotonScalarScalar}, it is easy to check that the delta-function $\delta(\omega_i)$ forces $\Delta_i=1$ with $i=2,3$. In the rest of this section, we will focus on the celestial amplitude $\mathcal{A}^{\Delta_i}_{1^a\rightarrow 2_{\text{soft}}^b3^c}$ with $\Delta_2=1$. The celestial amplitude with $\Delta_3=1$ can be derived in a similar way.

Due to $\Delta_2=1$, we should use $A^{\text{log},-}_{\mu z}$ to compute  $\mathcal{A}^{\Delta_i}_{1^a\rightarrow 2_{\text{soft}}^b3^c}$. In other words, we have 
\begin{align}
\begin{split}
\mathcal{A}^{\Delta_i}_{1^a\rightarrow 2_{\text{soft}}^b3^c}=&(2\pi)^4\int\frac{d^3q^{\prime}_1}{q^{\prime0}_1}\frac{d^3q^{\prime}_2}{q^{\prime0}_2}\frac{d^3q^{\prime}_3}{q^{\prime0}_3}A^{\Delta_1,+}_{\mu a}(\hat{q}_1;q^{\prime}_1)A^{\text{log},-}_{\nu b}(\hat{q}_2;q^{\prime}_2)\widetilde{A}^{\Delta_3,-}_{\rho c}(\hat{q}_3;q^{\prime}_3)\\
&\times\delta^{(4)}(q^{\prime}_1-q^{\prime}_2-q^{\prime}_3)\bigg[g^{\mu\nu}(q^{\prime}_1+q^{\prime}_2)^{\rho}+g^{\nu\rho}(-q^{\prime}_2+q^{\prime}_3)^{\mu}+g^{\rho\mu}(-q^{\prime}_3-q^{\prime}_1)^{\nu}\bigg]\;,
\end{split}
\end{align}
where $A^{\Delta,\pm}_{\mu a}(\hat{q};q^{\prime})$, $\widetilde A^{\Delta,\pm}_{\mu a}(\hat{q};q^{\prime})$, and $A^{\text{log},\pm}_{\mu a}(\hat{q};q^{\prime})$ are the Fourier transforms of \eqref{eq:SpinOne}, \eqref{eq:At}, and \eqref{eq:Alog}.
Equipped with the momentum conservation and the fact that $A^{\Delta_i,+}_{\mu \bar{z}}(\hat{q}_i;q^{\prime}_i)$ and $\widetilde{A}^{\Delta_i,-}_{\rho\bar{z}}(\hat{q}_i;q^{\prime}_i)$ are transverse with respect to the momentum $q^{\prime}_i$, we can write $\mathcal{A}^{\Delta_i}_{1^a\rightarrow 2_{\text{soft}}^b3^c}$ as
\begin{align}
\begin{split}
\mathcal{A}^{\Delta_i}_{1^a\rightarrow 2_{\text{soft}}^b3^c}=&(2\pi)^4\int\frac{d^3q^{\prime}_1}{q^{\prime0}_1}\frac{d^3q^{\prime}_2}{q^{\prime0}_2}\frac{d^3q^{\prime}_3}{q^{\prime0}_3}A^{\Delta_1,+}_{\mu a}(\hat{q}_1;q^{\prime}_1)A^{\text{log},-}_{\nu b}(\hat{q}_2;q^{\prime}_2)\widetilde{A}^{\Delta_3,-}_{\rho c}(\hat{q}_3;q^{\prime}_3)\\
&\times\delta^{(4)}(q^{\prime}_1-q^{\prime}_2-q^{\prime}_3)\bigg[2g^{\mu\nu}q^{\prime\rho}_1+2g^{\nu\rho}q^{\prime\mu}_3+g^{\rho\mu}(-q^{\prime}_3-q^{\prime}_1)^{\nu}\bigg]\;.
\end{split}
\end{align}
Following the steps in Section \ref{sec:PhotonScalarScalar}, $\mathcal{A}^{\Delta_i}_{1^a\rightarrow 2_{\text{soft}}^b3^c}$ can be recast into the form as
\begin{align}
\begin{split}
\mathcal{A}^{\Delta_i}_{1^a\rightarrow 2_{\text{soft}}^b3^c}=&i\int d^4X\bigg(-2\widetilde{A}^{\Delta_3,-}_{3,\rho c}\partial^{\rho}A^{\Delta_1,+}_{1,\mu a}A^{\text{log},-}_{2,\nu b}\eta^{\mu\nu}+2A^{\Delta_1,+}_{1,\mu a}A^{\text{log},-}_{2,\nu b}\partial^{\mu}\widetilde{A}^{\Delta_3,-}_{3,\rho c}\eta^{\rho\nu}\\
&\qquad+A^{\text{log},-}_{2,\nu b}\bigg[\partial^{\nu}A^{\Delta_1,+}_{1,\mu a}\widetilde{A}^{\Delta_3,-}_{3,\rho c}-A^{\Delta_1,+}_{1,\mu a}\partial^{\nu}\widetilde{A}^{\Delta_3,-}_{3,\rho c}\bigg]\eta^{\mu\rho}\bigg)\;.
\end{split}
\end{align}
Integrating by parts and using the  fact that $\partial^{\rho}\widetilde{A}^{\Delta_3,-}_{3,\rho c}=\partial^{\mu}A^{\Delta_1,+}_{1,\mu a}=0$, one can show that the pure gauge in $A^{\text{log},-}_{2,\nu b}$ does not contribute to $\mathcal{A}^{\Delta_2=1}_{1^a\rightarrow 2^b3^c}$. This leads to
\begin{align}
\begin{split}
\mathcal{A}^{\Delta_i}_{1^a\rightarrow 2_{\text{soft}}^b3^c}=&i\int d^4X\bigg(-2\widetilde{A}^{\Delta_3,-}_{3,\rho c}\partial^{\rho}A^{\Delta_1,+}_{1,\mu a}V^{\text{log},-}_{2,\nu b}\eta^{\mu\nu}+2A^{\Delta_1,+}_{1,\mu a}V^{\text{log},-}_{2,\nu b}\partial^{\mu}\widetilde{A}^{\Delta_3,-}_{3,\rho c}\eta^{\rho\nu}\\
&\qquad+V^{\text{log},-}_{2,\nu b}\bigg[\partial^{\nu}A^{\Delta_1,+}_{1,\mu a}\widetilde{A}^{\Delta_3,-}_{3,\rho c}-A^{\Delta_1,+}_{1,\mu a}\partial^{\nu}\widetilde{A}^{\Delta_3,-}_{3,\rho c}\bigg]\eta^{\mu\rho}\bigg)\;.
\end{split}
\end{align}
Since $V^{\text{log},-}_{2,\nu b}\propto X_{\nu}$ and $X^{\mu}\widetilde{A}^{\Delta_3,-}_{3,\mu c}=X^{\mu}A^{\Delta_1,+}_{1,\mu a}=0$, we find that
\begin{align}
\begin{split}
&-2\widetilde{A}^{\Delta_3,-}_{3,\rho c}\partial^{\rho}A^{\Delta_1,+}_{1,\mu a}V^{\text{log},-}_{2,\nu b}\eta^{\mu\nu}+2A^{\Delta_1,+}_{1,\mu a}V^{\text{log},-}_{2,\nu b}\partial^{\mu}\widetilde{A}^{\Delta_3,-}_{3,\rho c}\eta^{\rho\nu}=0\;,
\end{split}
\end{align}
leading to
\begin{align}
\begin{split}
\mathcal{A}^{\Delta_i}_{1^a\rightarrow 2_{\text{soft}}^b3^c}=&i\int d^4X V^{\text{log},-}_{2,\nu b}\bigg[\partial^{\nu}A^{\Delta_1,+}_{1,\mu a}\widetilde{A}^{\Delta_3,-}_{3,\rho c}-A^{\Delta_1,+}_{1,\mu a}\partial^{\nu}\widetilde{A}^{\Delta_3,-}_{3,\rho c}\bigg]\eta^{\mu\rho}\;.
\end{split}
\end{align}
Using \eqref{eq:SpinOne}, \eqref{eq:At} and \eqref{eq:Alog} we get
\begin{align}\label{eq:GluonGluonGluon}
\begin{split}
\mathcal{A}^{\Delta_i}_{1^a\rightarrow 2_{\text{soft}}^b3^c}=&\frac{2i(\Delta_1+\Delta_3-2)}{N_{\Delta_1}^+N_{2-\Delta_3}^{-}}\lim_{\Delta_2^{\prime}\rightarrow0}\frac{1}{N_{2-\Delta_2^{\prime}}^{-}\Delta_2^{\prime}}\bigg(\frac{(\hat{q}_1\cdot\partial_{3c}\hat{q}_3)}{\Delta_1}\partial_{1a}\partial_{2b}+(\partial_{1a}\hat{q}_1\cdot\partial_{3c}\hat{q}_3)\partial_{2b}
\\
&\qquad+\frac{(\hat{q}_1\cdot\hat{q}_3)}{\Delta_1\Delta_3}\partial_{1a}\partial_{2b}\partial_{3c}+\frac{(\partial_{1a}\hat{q}_1\cdot\hat{q}_3)}{\Delta_3}\partial_{2b}\partial_{3c}\bigg)\mathcal{A}^{\Delta_3\Delta^{\prime}_2\Delta_1}_{1^0\rightarrow2^03^0}\;.
\end{split}
\end{align}

Without loss of generality, we take $b=-$. Then, there are four possible helicity configurations: $(-,-,+)$, $(+,-,-)$, $(-,-,-)$, and $(+,-,+)$. Using \eqref{eq:GluonGluonGluon}, one can show that the celestial amplitudes $\mathcal{A}^{\Delta_i}_{1^-\rightarrow 2_{\text{soft}}^-3^-}$ and $\mathcal{A}^{\Delta_i}_{1^+\rightarrow 2_{\text{soft}}^-3^+}$ vanish. This is consistent with the result from standard CFT. Three-point conformal correlation function involving one conserved current vanishes unless the remaining two operators have the same weight. On the other hand, celestial amplitudes $\mathcal{A}^{\Delta_i}_{1^-\rightarrow 2_{\text{soft}}^-3^+}$ and $\mathcal{A}^{\Delta_i}_{1^+\rightarrow 2_{\text{soft}}^-3^-}$ do not vanish and take the form as standard conformal three-point function with one conserved current, \textit{i.e.,} we have 
\begin{align}\label{eq:MinusPlusMinus}
\begin{split}
\mathcal{A}^{\Delta_i}_{1^-\rightarrow 2_{\text{soft}}^-3^+}=&\frac{4i(2\pi)^3(\Delta_1-1)\sin(\Delta_1\pi)e^{i\Delta_1\pi}}{\Delta_1}(\frac{1}{z_{12}}+\frac{1}{z_{23}})\frac{1}{z_{13}^{\Delta_1-1}\bar{z}_{13}^{\Delta_1+1}}\delta(\Delta_1-\Delta_3)\;,
\end{split}
\end{align}
where the conformal weights are $(\frac{\Delta_1-1}{2},\frac{\Delta_1+1}{2})$, $(1,0)$ and $(\frac{\Delta_3-1}{2},\frac{\Delta_3+1}{2})$, and
\begin{align}\label{eq:PlusPlusPlus}
\begin{split}
\mathcal{A}^{\Delta_i}_{1^+\rightarrow 2_{\text{soft}}^-3^-}=&\frac{4i(2\pi)^3(\Delta_1-1)\sin(\Delta_1\pi)e^{i\Delta_1\pi}}{\Delta_1}(\frac{1}{z_{12}}+\frac{1}{z_{23}})\frac{1}{z_{13}^{\Delta_1+1}\bar{z}_{13}^{\Delta_1-1}}\delta(\Delta_1-\Delta_3)\;,
\end{split}
\end{align}
where the conformal weights are $(\frac{\Delta_1+1}{2},\frac{\Delta_1-1}{2})$, $(1,0)$ and $(\frac{\Delta_3+1}{2},\frac{\Delta_3-1}{2})$. \eqref{eq:MinusPlusMinus} and \eqref{eq:PlusPlusPlus} indicate that the soft gluon with helicity $+$ is mapped to be a holomorphic current $j_{z}$ with conformal dimension one. The OPE between a soft gluon and another gluon takes the form as
\begin{align}\label{eq:GluonOPE}
\begin{split}
j_{z}(z_2)\mathcal{O}^{\Delta_1,\pm}_{z}(z_1)\sim&\frac{4i(2\pi)^3(\Delta_3-1)\sin(\Delta_3\pi)e^{i\Delta_3\pi}\delta(\Delta_1-\Delta_3)}{\Delta_3C^{\Delta_1\Delta_3}_{zz}}\frac{1}{z_{12}}\mathcal{O}^{\Delta_3,\mp}_{z}(z_2)\;,
\end{split}
\end{align}
where $C^{\Delta_1\Delta_3}_{zz}$ is the coefficient of the two-point celestial amplitude of gluons, which contains a delta function $\delta(\Delta_1-\Delta_3)$ canceling the delta function in the numerator.

\subsection{Comparing with the results in Klein space }

The celestial amplitude of three gluons in Klein space was computed in \cite{Pasterski:2017ylz} by performing Mellin transformation on MHV amplitude. To compare our results with their, we first derive the MHV amplitude in Klein space from the scattering amplitude \eqref{eq:MGluonGluonGluon}. Without loss of generality, we take the helicities as $(-,-,+)$. The MHV amplitude can be obtained from $\mathcal{M}^{\mu\nu\rho}$ by contracting with the polarization vectors $\partial_{\bar{z}_1}\hat{q}_{1\mu}\partial_{\bar{z}_2}\hat{q}_{2\nu}\partial_{z_3}\hat{q}_{3\rho}$, leading to
\begin{align}
\begin{split}
\mathcal{M}_{--+}=&\partial_{\bar{z}_1}\hat{q}_{1\mu}\partial_{\bar{z}_2}\hat{q}_{2\nu}\partial_{z_3}\hat{q}_{3\rho}\mathcal{M}^{\mu\nu\rho}(q_1,q_2,q_3)\\
=&2(2\pi^4)\delta^{(4)}(q_1-q_2-q_3)\bigg[\partial_{\bar{z}_1}\hat{q}_1\cdot(-q_2+q_3)+\partial_{\bar{z}_2}\hat{q}_2\cdot(-q_3-q_1)\bigg]\;,
\end{split}
\end{align}
where we used the fact $\partial_{\bar{z}_1}\hat{q}_1\cdot\partial_{\bar{z}_2}\hat{q}_2=0$ and $\partial_{z_1}\hat{p}_1\cdot\partial_{\bar{z}_3}\hat{p}_3=2$. With the help of the momentum conservation and the fact that polarization vector $\epsilon_{i\pm}$ is orthogonal to the momentum $q_{i}$, we get
\begin{align}
\begin{split}
\mathcal{M}_{--+}=&2(2\pi^4)\delta^{(4)}(q_1-q_2-q_3)\bigg[-2\omega_2\partial_{\bar{z}_1}\hat{q}_1\cdot\hat{q}_2-2\omega_1\hat{q}_1\cdot\partial_{\bar{z}_2}\hat{q}_2\bigg]\;.
\end{split}
\end{align}
The support of delta-function $\delta^{(4)}(q_1-q_2-q_3)$ contains soft region and colinear region in both Minkowski space and Klein space. The colinear region demands that
\begin{align}\label{eq:Colinear}
\hat{q}_1\cdot\hat{q}_2=\hat{q}_1\cdot\hat{q}_3=0\;.
\end{align}
In Minkowski space, \eqref{eq:Colinear} implies $z_1=z_2=z_3$ and $\bar{z}_1=\bar{z}_2=\bar{z}_3$, leading to vanishing $\mathcal{M}_{--+}$. However, since $z_i$ and $\bar{z}_i$ are independent real variables in Klein space, \eqref{eq:Colinear} implies $z_1=z_2=z_3$ or $\bar{z}_1=\bar{z}_2=\bar{z}_3$. Indeed, the momentum conservation in Klein space has support at \footnote{Since $\omega_i\geq0$, we must have $z_3<z_1<z_2$ or $z_3>z_1>z_2$.}
\begin{align}
\bar{z}_1=\bar{z}_2=\bar{z}_3\;,\qquad\omega_1=-\frac{z_{23}}{z_{12}}\omega_3\;,\qquad\omega_2=-\frac{z_{13}}{z_{12}}\omega_3\;,
\end{align}
which corresponds to the colinear region. On this support, we find that
\begin{align}\label{eqn:MHV_Klein}
\mathcal{M}_{--+}=8(2\pi)^4\delta^{(4)}(q_1-q_2-q_3)\frac{\omega_1\omega_2}{\omega_3}\frac{z_{12}^3}{z_{23}z_{31}}\;,
\end{align}
which agrees with the MHV amplitude up to a constant.
This leads to the following celestial amplitude in Klein space \cite{Pasterski:2017ylz}
\begin{align}\label{eq:GluonKlein}
\mathcal{A}_{--+}=-\pi\delta(\sum_i\lambda_i)\frac{\text{sgn}(z_{12}z_{23}z_{31})\delta(\bar{z}_{12})\delta(\bar{z}_{13})}{|z_{12}|^{-1-i\lambda_3}|z_{23}|^{1-i\lambda_1}|z_{13}|^{1-i\lambda_2}}\;,   
\end{align}
which apparently takes a very different form than our results \eqref{eq:MinusPlusMinus} and \eqref{eq:PlusPlusPlus}. As we have seen above, \eqref{eqn:MHV_Klein} only includes the contribution from the colinear region.
Therefore, the celestial amplitude \eqref{eq:GluonKlein} also only counts the contribution from the colinear region. This statement can also be seen from the delta-function $\delta(\bar{z}_{12})\delta(\bar{z}_{13})$ in \eqref{eq:GluonKlein}. 

We conclude that the celestial amplitude of three gluons in Minkowski space only receives contributions from the soft region, while the celestial amplitude of three gluons in Klein space receives contributions from both the soft and colinear regions. To complete the results in Klein space, one still needs to add the contribution from the soft region to the results in \eqref{eq:GluonKlein}. This soft contribution in Klein space can be obtained in a similar way that we did in Minkowski space.

\section{Discussions}\label{sec:discussion}

 In this paper, we gave the first computation of the celestial amplitudes of three massless particles in Minkowski space. We showed that due to the existence of soft and colinear regions in the support of the delta-function for momentum conservation, the celestial amplitudes of three massless particles in Minkowski space do not vanish and take the standard form of correlation functions in CFT. Focusing on the specific amplitudes of scalar-scalar-scalar, scalar-photon-scalar, scalar-graviton-scalar, and gluon-gluon-gluon, we found that the massless three-point celestial amplitudes of scalars receive contributions from both the soft and colinear regions, while the massless three-point celestial amplitudes of gluon and graviton receive contribution only from the soft region. Moreover, by looking at the celestial amplitudes of scalar-photon-scalar and gluon-gluon-gluon, we found that the scattering amplitudes involving a soft spin-one particle are mapped to be conformal correlators involving a spin-one current on the celestial sphere. These two examples directly confirmed the relation between soft spin-one particles in Minkowski space and conserved currents on the celestial sphere at the level of three-point amplitudes. In addition, we also found that the soft graviton with positive helicity in the scalar-graviton-scalar scattering is mapped to be a primary operator with conformal weight $(\frac{3}{2},-\frac{1}{2})$. By further taking a $\bar z$-derivative on this primary operator, we obtained the supertranslation current. We derived the OPE between the supertranslation current and scalar primaries and found the OPE matches with the one found in \cite{Donnay:2018neh}.

Several interesting open questions ensue from our work. First, 
it would be interesting to extend our analysis to higher-point celestial amplitudes to further examine the relation between soft spinning particles and conserved currents.
Similar to the three-point case, we expect our prescription \eqref{eq:CA} for the celestial amplitudes and the logarithmic conformal primary wave functions \eqref{eq:Alog_def} and \eqref{eq:hlog} should play an important role in the higher-point case.
While only the soft region has contributed in the three-point celestial amplitudes with spinning particles, the higher-point celestial amplitudes receive contributions from other solutions to the momentum conservation. Thus, to study the conserved currents in higher-point celestial amplitudes, one must extract their contributions from taking the (conformally) soft limit or projecting the wave function onto the logarithmic conformal primary wave functions.

Another avenue would be to explore how the stress tensor emerges from the soft region at the level of three-point celestial amplitudes. The celestial stress tensor is usually constructed as a shadow of the subleading conformally soft graviton \cite{Cachazo:2014fwa,Kapec:2016jld,Kapec:2014opa}. The obstruction of this way in the three-point celestial amplitude is that the energy $\omega$ is strictly equal to zero. Then it is unclear how to obtain the subleading conformally soft graviton.

Finally, it would be of great interest to study the celestial amplitudes of three conserved spin-one currents as well as of three stress tensors. These two correlators are important as they encode the information of the level and the central charge. To get these two correlators, one must take double or triple (conformally) soft limits of celestial amplitudes. The double soft limit which is necessary to consider $TT$ OPE was studied in \cite{Fotopoulos:2020bqj}. However, there is an obstruction in \cite{Fotopoulos:2020bqj} to reproducing the $TT$ OPE by taking the double soft limit. \footnote{Multiple soft insertion was studied in \cite{Ball:2022bgg}.} A possible resolution to this obstruction was proposed very recently in \cite{Banerjee:2022wht} for amplitudes in the Klein space, which used the ambidextrous basis \cite{Sharma:2021gcz,Jorge-Diaz:2022dmy}. Their basis involves the light transformation, and hence cannot be defined in Minkowski space, which has an Euclidean celestial sphere.
It would be interesting to develop techniques in Minkowski space that allow us to take the double soft limit. Perhaps, one can naively use the shadow transformation to replace the light transformation. \footnote{A correspondence between Fourier transforms and shadow transforms was studied in \cite{Brown:2022miw}. }

\section*{Acknowledgements}

We owe our gratitude to Andrew Strominger for a correspondence that partially motivated this work.
CC is partly supported by National Key R\&D Program of China (NO. 2020YFA0713000).

\appendix

\section{Conformal primary basis for general spin}\label{app:CPB_general_spin}

In this appendix, we give a covariant formalism that produces the massless and massive conformal primary basis given in Section \ref{sec:CPB_general_spin}.

\subsection{Massless conformal primary basis}

We note that the little group for massless particles in $\mathbb{R}^{1,3}$ is exactly the rotation subgroup inside the $2$-dimensional conformal group. Thus, if we treat the four-dimensional Minkowski space as the embedding space and the two-dimensional celestial sphere as the position space, conformal primary wave functions $\Phi^{\Delta,\pm}_{\{\mu\}\{a\}}(\hat{q};X)$ for massless particles in the little group representation $\mathbf{N}$ can be built from the embedding space conformal two-point function $\langle\mathcal{O}_{\Delta}^{\mathbf{N}}(\hat{q})\mathcal{O}_{\Delta^{\prime}}^{\mathbf{N}}(\hat{q}^{\prime})\rangle$ through \footnote{The embedding space formalism used in this paper can be found in \cite{Weinberg:2010fx,Weinberg:2012mz,Rychkov:2016iqz} and the references therein. There is a different CFT embedding space formalism which was studied in \cite{Fortin:2019dnq,Skiba:2019cmf}.}
\begin{align}\label{eq:MasslessCPB}
\begin{split}
\Phi^{\Delta,\pm}_{\{\mu\}\{a\}}(\hat{q};X)=&\mathcal{N}_{\Delta,\ell}\int\frac{d^{3}q^{\prime}}{q^{0\prime}}e^{\pm iq^{\prime}\cdot X}{\Pi(X,q^{\prime})_{\{\mu\}}}^{\{\nu\}}\langle\mathcal{O}^{\mathbf{N}}_{\Delta^{\prime}\{\nu\}}(q^{\prime})(\mathcal{P}\mathcal{O}^{\mathbf{N}})_{\Delta\{a\}}(\hat{q})\rangle\;,
\end{split}
\end{align}
where $\ell$ denotes the spin and the normalization factor $\mathcal{N}_{\Delta,\ell}$ is defined in \eqref{eq:NDelta}.
Explanations are made here. $\mu,\nu,\dots$ and $a,b,\dots$ were used to denote indices in $\mathbb{R}^{1,3}$ and $\mathbb{R}^2$, respectively. $\mathcal{P}$ was introduced to projects the embedding space operator $\mathcal{O}^{\mathbf{N}}_{\Delta\{\mu\}}(\hat{q})$ into the position space conformal primary operator and ${\Pi(X,q^{\prime} )_{\{\mu\}}}^{\{\nu\}}$ was introduced to make the conformal primary basis satisfy the equation of motion. We also note that the projector $\Pi(X,q)_{\{\mu\}}^{\phantom{\{\mu\}}\{\nu\}}$ must transform as an element in the irreducible representation $\mathbf{N}$ 
\begin{align}\label{eq:PiTransform}
\Pi(\Lambda X,\Lambda q)=R^{\mathbf{N}}(\Lambda)\Pi(X,q)\left(R^{\mathbf{N}}\right)^{-1}(\Lambda)\;,   
\end{align}
where ${R^{\mathbf{N}}(\Lambda)_{\{\mu\}}}^{\{\nu\}}$ is the matrix element implementing the action of $\Lambda$ in the Lorentz group $SO(1,3)$.

Next, we prove that \eqref{eq:MasslessCPB} is conformally covariant. The conformal covariance of \eqref{eq:MasslessCPB} is ensured by the insertion of the conformal two-point function $\langle\mathcal{O}^{\mathbf{N}}_{\Delta^{\prime}\{\nu\}}(\hat{q}^{\prime})(\mathcal{P}\mathcal{O}^{\mathbf{N}})_{\Delta\{a\}}(\hat{q})\rangle$. Indeed, under the Lorentz transformation $X\rightarrow \Lambda X$ and $\hat{q}\rightarrow\hat{q}^{\prime}$, we have \footnote{To simplify the notation, we omit the labeling $\mathbf{N}$ in the rest of this paper.}
\begin{align}
\begin{split}
\Phi^{\Delta,\pm}(\hat{q}^{\prime};\Lambda X)=&\int\frac{d^{3}q^{\prime\prime}}{2q^{0\prime\prime}}e^{\pm iq^{\prime\prime}\cdot(\Lambda X)}\Pi(\Lambda X,q^{\prime\prime})\cdot\langle\mathcal{O}_{\Delta^{\prime}}(q^{\prime\prime})(\mathcal{P}\mathcal{O}_{\Delta})(\hat{q}^{\prime})\rangle\;.
\end{split}
\end{align}
Changing $q^{\prime\prime}$ to $\Lambda q^{\prime\prime}$ and using the fact that $(\Lambda q^{\prime\prime})\cdot(\Lambda X)=q^{\prime\prime}\cdot X$ as well as $\hat{q}^{\prime}=\Lambda \hat{q}$ leads to
\begin{align}
\begin{split}
      \Phi_{\Delta,\pm}(\hat{q}^{\prime};\Lambda X)=&\int\frac{d^{3}q^{\prime\prime}}{2q^{0\prime\prime}}e^{\pm iq^{\prime\prime}\cdot X}\Pi(\Lambda X,\Lambda q^{\prime\prime})\cdot\langle\mathcal{O}_{\Delta^{\prime}}(\Lambda q^{\prime\prime})(\mathcal{P}\mathcal{O}_{\Delta})(\Lambda\hat{q})\rangle\;.
\end{split}
\end{align}
Using \eqref{eq:PiTransform} and the fact that
\begin{align}
\begin{split}
\langle\mathcal{O}_{\Delta^{\prime}}(\Lambda q^{\prime\prime})(\mathcal{P}\mathcal{O}_{\Delta})(\Lambda\hat{q})\rangle=\langle R(\Lambda)\mathcal{O}_{\Delta^{\prime}}( q^{\prime\prime})(\mathcal{P}R(\Lambda)\mathcal{O}_{\Delta})(\hat{q})\rangle\;,  
\end{split}
\end{align}
we then get
\begin{align}
\begin{split}
      \Phi_{\Delta,\pm}(\hat{q}^{\prime};\Lambda X)=&R(\Lambda)\int\frac{d^{3}q^{\prime\prime}}{2q^{0\prime\prime}}e^{\pm iq^{\prime\prime}\cdot X}\Pi(X, q^{\prime\prime})\cdot\langle\mathcal{O}_{\Delta^{\prime}}(q^{\prime\prime})(\mathcal{P}R(\Lambda)\mathcal{O}_{\Delta})(\hat{q})\rangle\;.
\end{split}
\end{align}
Since $(\mathcal{P}\mathcal{O}_{\Delta})(\hat{q})$ projects $\mathcal{O}_{\Delta}(\hat{q})$ into a conformal primary operator in two-dimensional space, the following identity must hold
\begin{align}
\mathcal{P}(\mathcal{R}(\Lambda)\mathcal{O}_{\Delta})(\hat{q})=\left|\frac{\partial z^{\prime}}{\partial z}\right|^{-\frac{\Delta}{2}}D(\Lambda)(\mathcal{P} \mathcal{O}_{\Delta})(\hat{q})\;,
\end{align}
where ${D(\Lambda)_{\{a\}}}^{\{b\}}$ is the matrix element implementing the action of Lorentz group $SO(2)$. This leads to
\begin{align}
\begin{split}
      \Phi^{\Delta,\pm}_{\{\mu\}\{a\}}(\hat{q}^{\prime};\Lambda X)=&R(\Lambda)D(\Lambda)\left|\frac{\partial z^{\prime}}{\partial z}\right|^{-\frac{\Delta}{2}}\int\frac{d^{3}q^{\prime\prime}}{2q^{0\prime\prime}}e^{\pm iq^{\prime\prime}\cdot X}\Pi(X, q^{\prime\prime})\cdot\langle\mathcal{O}_{\Delta^{\prime}}(q^{\prime\prime})(\mathcal{P}\mathcal{O}_{\Delta})(\hat{q})\rangle\\
      =&{R(\Lambda)_{\{\mu\}}}^{\{\nu\}}{D(\Lambda)_{\{a\}}}^{\{b\}}\left|\frac{\partial z^{\prime}}{\partial z}\right|^{-\frac{\Delta}{2}}\Phi^{\Delta,\pm}_{\{\mu\}\{a\}}(\hat{q};X)\;,
\end{split}
\end{align}
which verify the transformation property.

Let us focus on massless spin-$\ell$ particles. The conformal primary wave function \eqref{eq:MasslessCPB} should satisfy the Fronsdal equation,
\begin{align}
\Box \Phi^{\Delta,\pm}_{\mu_1\cdots\mu_\ell,\{a^\ell\}}-\ell \partial_{(\mu_1}\partial^\nu \Phi^{\Delta,\pm}_{\mu_2\cdots\mu_\ell)\nu,\{a^\ell\}}+\frac{\ell(\ell-1)}{4\ell}\eta_{(\mu_1\mu_2}\partial^{\nu_1}\partial^{\nu_2}\Phi^{\Delta,\pm}_{\mu_3\cdots\mu_\ell)\nu_1\nu_2,\{a^\ell\}}=0\,.
\end{align}
After imposing the Lorentz gauge $\partial^\nu\Phi^{\Delta,\pm}_{\nu\mu_2\cdots\mu_\ell,\{a^\ell\}}=0$, the Fronsdal equation becomes the massless Klein-Gordon equation, which constrains
\begin{align}\label{eqn:spin-l_Pi}
\Pi(q',X)_{\{\mu^\ell\}}{}^{\{\nu^\ell\}}=\delta_{\{\mu^\ell\}}{}^{\{\nu^\ell\}}\,.
\end{align}
Furthermore, the two-point conformal correlation function of spin-$\ell$ operators takes the form as 
\begin{align}\label{eqn:spin-l_OO}
\langle\mathcal{O}_{\Delta}^{\{\nu^{\ell}\}}(\hat{q}^{\prime})\mathcal{O}_{\Delta;\{\mu^{\ell}\}}(\hat{q})\rangle=\frac{{(\mathcal{P}_I^{\ell})_{\{\mu^{\ell}\}}}^{\{\nu^{\ell}\}}(\hat{q},\hat{q}^{\prime})}{(-\frac{1}{2}\hat{q}^{\prime}\cdot\hat{q})^{\Delta}}\;.
\end{align}
Plugging \eqref{eqn:spin-l_Pi} and \eqref{eqn:spin-l_OO} into \eqref{eq:MasslessCPB}, we find the formula \eqref{eq:MasslessSpinl}.

\subsubsection*{Relation to shadow conformal primary basis}\label{app:SCPB}

In this appendix, we will specialize the formula \eqref{eq:MasslessSpinl} to the $\ell=0$ and $1$, and show that they are proportional to the shadow conformal primary basis $\widetilde\phi^{\pm}_{\Delta}$ and $\widetilde{\mathcal{A}}^{\Delta,\pm}_{\mu a}$, respectively.

First, for $\ell=0$, it is straightforward to see that
\begin{align}
\begin{split}
\Phi^{\Delta,\pm}(\hat{q};X)=&\mathcal{N}_{\Delta,0}\int\frac{d^{3}q^{\prime}}{q^{0\prime}}\frac{1}{(-\hat{q}^{\prime}\cdot\hat{q})^{\Delta}}e^{\pm iq^{\prime}\cdot X}
=2^{1-\Delta}\widetilde\phi^{\pm}_{\Delta}(z;X)\;,
\end{split}
\end{align}
where $\widetilde\phi^{\pm}_{\Delta}(z;X)$ is the shadow conformal primary wave function \eqref{eq:phit=phi}.

Next, for $\ell=1$, we have
\begin{align}
\begin{split}
\Phi^{\Delta,\pm}_{\mu a}(\hat{q};X)=&
\mathcal{N}_{\Delta,1}\partial_a\hat{q}^{\nu}\int\frac{d^{3}q^{\prime}}{q^{0\prime}}\frac{I_{\mu\nu}(\hat{q},\hat{q}^{\prime})}{(-\hat{q}^{\prime}\cdot\hat{q})^{\Delta}}e^{\pm iq^{\prime}\cdot X}
\\
=&2\mathcal{N}_{\Delta,1}\partial_a\hat{q}^{\nu}\int D^2\hat{q}^{\prime}\int_0^{\infty}d\omega\,\omega^{1-\Delta}\frac{(-\hat{q}\cdot \hat{q}^{\prime})\eta_{\mu\nu}+\hat{q}_{\mu}\hat{q}^{\prime}_{\nu}}{(-\hat{q}\cdot \hat{q}^{\prime})^{\Delta+1}}e^{\pm i\omega\hat{q}^{\prime}\cdot X}\;,
\end{split}
\end{align}
where we defined $D^2\hat{q}^{\prime}\equiv d^2z^{\prime}$. The term containing $\eta_{\mu\nu}$ is proportional to the scalar shadow conformal primary basis
\begin{align}
\begin{split}
&2\partial_a\hat{q}^{\nu}\int D^2\hat{q}^{\prime}\int_0^{\infty}d\omega\,\omega^{1-\Delta}\frac{(-\hat{q}\cdot \hat{q}^{\prime})\eta_{\mu\nu}}{(-\hat{q}\cdot \hat{q}^{\prime})^{\Delta+1}}e^{\pm i\omega\hat{q}^{\prime}\cdot X}=\frac{2^{2-\Delta}\pi(\mp i)^{2-\Delta}\Gamma[1-\Delta](-X^2)^{\Delta-1}}{(-\hat{q}\cdot X)^{\Delta}}\;.
\end{split}
\end{align}
To compute the remaining part, we note that the Lorentz invariance and scaling behaviour fix that
\begin{align}
2\int D^2\hat{q}^{\prime}\int_0^{\infty}d\omega\,\omega^{1-\Delta}\frac{\hat{q}^{\prime}_{\nu}}{(-\hat{q}\cdot \hat{q}^{\prime})^{\Delta+1}}e^{\pm i\omega\hat{q}^{\prime}\cdot X}=N\frac{X_{\nu}(-X^2)^{\Delta-1}}{(-\hat{q}\cdot X)^{\Delta+1}}+\text{$\hat{q}_{\nu}$-term}\;.
\end{align}
To determine the proportional constant $N$, we contract both sides with $-\hat{q}^{\nu}$, giving
\begin{align}
N=2^{2-\Delta}\pi(\mp i)^{2-\Delta}\Gamma[1-\Delta]\;.
\end{align}
We find that $\Phi^{\Delta,\pm}_{\mu a}(X,z)$ is propotional to $\widetilde{A}^{\Delta,\pm}_{\mu a}$,
\begin{align}\label{eq:Phi=At}
\begin{split}
\Phi^{\Delta,\pm}_{\mu a}(\hat{q};X)=&2^{2-\Delta}\pi(\mp i)^{2-\Delta}\Gamma[1-\Delta]\mathcal{N}_{\Delta,1}\frac{(-X^2)^{\Delta-1}}{(-\hat{q}\cdot X\mp i\epsilon)^{\Delta+1}}\bigg((-\hat{q}\cdot X)\partial_a\hat{q}_{\mu}+\hat{q}_{\mu}\partial_a\hat{q}\cdot X\bigg)\\
=&2^{1-\Delta}(\mp i)^{2-\Delta}\frac{\Gamma[3-\Delta]}{1-\Delta}\widetilde{A}^{\Delta,\pm}_{\mu a}(\hat{q};X)\;.
\end{split}
\end{align}
%

\subsubsection*{Mellin basis and light ray basis}

We note that besides the standard power-law form, the scalar two-point conformal correlations can also be a delta-function when $\Delta^{\prime}+\Delta=2$, \textit{i.e.,}
\begin{align}
\langle\mathcal{O}_{2-\Delta}(\hat{q}^{\prime})\mathcal{O}_{\Delta}(\hat{q})\rangle=\delta^{(2)}(z^{\prime}-z),
\end{align}
leading to
\begin{align}
\Phi^{\Delta,\pm}(\hat{q};X)=&2\int_0^{\infty}d\omega\,\omega^{\Delta-1}e^{\pm i\omega\hat{q}\cdot X}\;,
\end{align}
which is exactly the massless scalar conformal primary wave function given in \eqref{eq:Massless}.

Before going to spinning conformal primary basis, we mention that the formula \eqref{eq:MasslessCPB} can also be applied to Klein space. In Klein space, since $z$ and $\bar{z}$ become two independent real variables, the correlation functions in two-dimensional Lorentzian CFT can be written as a product of correlation functions in two one-dimensional CFTs. In this case, we can choose one of the two one-dimensional correlation functions to be one-dimensional delta function and another to be the power-law two-point function:
\begin{align}
\langle\mathcal{O}_{h,1-\bar{h}}(\hat{q}^{\prime})\mathcal{O}_{h,\bar{h}}(\hat{q})\rangle=\delta(\bar{z}^{\prime}-\bar{z})\frac{1}{(z^{\prime}-z)^{2h}}\;.
\end{align}
This leads to
\begin{align}\label{eq:LightRay}
\Phi^{\Delta,\pm}(\hat{q};X)=2^{1-\Delta}\int_{-\infty}^{\infty}dz^{\prime}\frac{1}{(z^{\prime}-z)^{2h}}\int_{0}^{+\infty}d\omega\omega^{-h+\bar{h}}e^{\pm i\omega\hat{q}(z^{\prime},\bar{z})\cdot X}\;.
\end{align}
Here we embed holomorphic coordinates $z$ and anti-holomorphic coordinates $\bar{z}$ into the on-shell momenta $q^{\mu}$ through 
\begin{align}
q^{\mu}=\omega\hat{q}^{\mu}=\omega(1+z\bar{z},z+\bar{z},z-\bar{z},1-z\bar{z})\;.
\end{align}
We note that \eqref{eq:LightRay} is the light transformation of the conformal primary basis \eqref{eq:Massless} \cite{Atanasov:2021cje}.

\subsection{Massive conformal primary basis}

With the help of spinning AdS bulk-to-boundary propogator, an index-free expression for massive spin-$\ell$ conformal primary basis has been obtained in \cite{Law:2020tsg}. In this subsection, we will using a different to construct the massive spin-$\ell$ conformal primary basis which satisfies the massive Klein-Gordon equation and transforms covariantly under $SL(2,\mathbb{C})$. Our conformal primary basis reproduces the results in \cite{Law:2020tsg}.

To construct the massive spin-$\ell$ conformal primary basis, we first note that massive spin-$\ell$ bosons are in the spin-$\ell$ representation of the little group $SO(3)$. This representaion can be labeled by two parameters $\ell$ and $J$ with $J=-\ell,-\ell+1,\cdots,\ell-1,\ell$ which are eigenvalues of the spin operators $\hat{S}^2$ and $\hat{S}_z$. On the other hand, the spin-$\ell$ operator of $SO(2)$ only has two components. To match the degree of freedoms, a massive spin-$\ell$ operator in Minkowski space should be mapped into a set of spin-$s$ operator $\mathcal{O}_s$ on the celestial sphere with $|s|=|J|$. 

To implement this map, we introduce the following two metrics built from the on-shell momenta $p$ and celestial coordinates $\hat{q}$:
\begin{align}
\begin{split}
&P^{\mu\nu}=g^{\mu\nu}+\hat{p}^{\mu}\hat{p}^{\nu}\;,\\
&Q^{\mu\nu}=g^{\mu\nu}-\frac{\hat{p}^{\mu}\hat{q}^{\nu}}{\hat{p}\cdot\hat{q}}-\frac{\hat{p}^{\nu}\hat{q}^{\mu}}{\hat{p}\cdot\hat{q}}-\frac{\hat{q}^{\mu}\hat{q}^{\nu}}{(\hat{p}\cdot\hat{q})^2}.
\end{split}
\end{align}
By construction, $P^{\mu\nu}$ and $Q^{\mu\nu}$ are symmetric with respect to $\mu$ and $\nu$ and satisfy the following transversality condition
\begin{align}\label{eq:TransC}
\hat{p}_{\mu}P^{\mu\nu}=\hat{p}_{\mu}Q^{\mu\nu}=\hat{q}_{\mu}Q^{\mu\nu}=0\;.
\end{align}
Using the transversality condition, it is easy to check that both $P^{\mu\nu}$ and $Q^{\mu\nu}$ are idempotent, \textit{i.e.,} 
\begin{align}
&P\cdot P=P\;,\qquad Q\cdot Q=Q.
\end{align}
Combining the idempotent with the fact that $\text{Tr}(P)=3$ and $\text{Tr}(Q)=2$, we conclude that $P^{\mu\nu}$ is a projector projecting a four-dimensional Lorentzian space into a three-dimensional subspace that is orthogonal to $\hat{p}^{\mu}$, while $Q^{\mu\nu}$ is a projector projecting a four-dimensional Lorentzian space into a two-dimensional subspace that is orthogonal to both $\hat{p}^{\mu}$ and $\hat{q}^{\mu}$. Since $P$ and $Q$ are metrics in three- and two-dimensional subspaces inside four-dimensional Lorentzian space, we can construct two spin-$\ell$ projection operator $\mathcal{P}^{\ell}_{P}$ and $\mathcal{P}^{\ell}_{Q}$ of subgroups $SO(3)$ and $SO(2)$ inside $SO(1,3)$ through simply replacing the metric $I$ in $\mathcal{P}^{\ell}_I$ by $P$ and $Q$, respectively.

Armed with $\mathcal{P}^{\ell}_{P}$ and $\mathcal{P}^{\ell}_{Q}$, the massive spin-$\ell$ conformal primary basis $\Phi^{\Delta,\ell,J,\pm}_{\{\mu^{\ell}\}\{a^|J|\};m}(\hat{q};X)$ takes the form as \eqref{eq:MassiveSpin} in Section \ref{sec:CPB_general_spin}. The explanation is made here. First, we insert the projection operator $\mathcal{P}^{\ell}_{P}$ to get spin-$\ell$ representation of the little group $SO(3)$. After that, we contract with $\ell-|J|$ $\hat{q}$s to get the representation with $|\hat{S}_z|=|J|$. Finally we contract with $\mathcal{P}^{|J|}_{Q}\cdot(\partial_z\hat{q})^{|J|}$ to get the spin-$|J|$ representation of $SO(2)$ on the celestial sphere. Since every step is conformally covariant, the conformal covariance of \eqref{eq:MassiveSpin} is manifested.

\section{Shadow celestial amplitude}\label{sec:ShadowAmplitude}

In this appendix, we compute the shadow celestial amplitudes $\widetilde{\mathcal{A}}^{\Delta_i}_{3^0\rightarrow1^02^0}$ of three massless scalars. After that, We perform the shadow transformations on $\widetilde{\mathcal{A}}^{\Delta_i}_{3^0\rightarrow1^02^0}$ to reproduce \eqref{eq:ScalarTwoOutgoing}. Moreover, armed with $\widetilde{\mathcal{A}}^{\Delta_i}_{3^0\rightarrow1^02^0}$, we also compute the shadow celestial amplitude $\widetilde{\mathcal{A}}^{\Delta_i}_{1^0\rightarrow2^+_{\text{soft}}3^0}$ involving one incoming massless scalar, one outgoing photon and one outgoing massless scalar. By performing the shadow transformation on $\widetilde{\mathcal{A}}^{\Delta_i}_{1^0\rightarrow2^+_{\text{soft}}3^0}$, we rederive \eqref{eq:PhotonScalarScalar}.

The shadow celestial amplitude $\widetilde{\mathcal{A}}^{\Delta_i}_{3^0\rightarrow1^02^0}$ of three massless scalars is given by
\begin{align}
\begin{split}
\widetilde{\mathcal{A}}^{\Delta_i}_{3^0\rightarrow 1^02^0}=&\frac{(2\pi)^4g\mathcal{N}_{\Delta_1,0}\mathcal{N}_{\Delta_2,0}\mathcal{N}_{\Delta_3,0}}{2^{3-\Delta_1-\Delta_2-\Delta_3}}\int\frac{d^{3}q^{\prime}_1}{q^{\prime0}_1}\frac{d^{3}q^{\prime}_2}{q^{\prime0}_2}\frac{d^{3}q^{\prime}_3}{q^{\prime0}_3} \frac{\delta^{(4)}(-q^{\prime}_1-q^{\prime}_2+q^{\prime}_3)}{(-\hat{q}_1\cdot q^{\prime}_1)^{\Delta_1}(-\hat{q}_2\cdot q^{\prime}_2)^{\Delta_2}(-\hat{q}_3\cdot q^{\prime}_3)^{\Delta_3}}\;.
\end{split}
\end{align}
Following the steps in Section \ref{sec:TwoOutgoing} and using the identity \cite{Dolan:2011dv}
\begin{align}\label{eq:triangle}
\begin{split}
\int d^2z_0\frac{1}{z_{01}^{a_1}z_{02}^{a_2}z_{03}^{a_3}}\frac{1}{\bar{z}_{01}^{\bar{a}_1}\bar{z}_{02}^{\bar{a}_2}\bar{z}_{03}^{\bar{a}_3}}=&\frac{2\pi\Gamma[1-a_1]\Gamma[1-a_2]\Gamma[1-a_3]}{\Gamma[\bar{a}_1]\Gamma[\bar{a}_2]\Gamma[\bar{a}_3]}\\
&\qquad\times\frac{1}{z_{12}^{1-a_3}z_{23}^{1-a_1}z_{31}^{1-a_2}}\frac{1}{\bar{z}_{12}^{1-\bar{a}_3}\bar{z}_{23}^{1-\bar{a}_1}\bar{z}_{31}^{1-\bar{a}_2}}\;.
\end{split}
\end{align}
which holds when $a_1+a_2+a_3=\bar{a}_1+\bar{a}_2+\bar{a}_3=2$, we get
\begin{align}\label{eq:ShadowAmplitudeScalar}
\widetilde{\mathcal{A}}_{3^0\rightarrow 1^02^0}^{\Delta_i}(z_i)=\frac{\widetilde{C}^{\Delta_i}_{3^0\rightarrow1^02^0}}{|z_{12}|^{\Delta_1+\Delta_2-\Delta_3}|z_{13}|^{\Delta_1-\Delta_2+\Delta_3}|z_{23}|^{\Delta_{2}+\Delta_3-\Delta_1}}\;,
\end{align}
where the coefficient $\widetilde{C}^{\Delta_i}_{3^0\rightarrow1^02^0}$ is
\begin{align}
\begin{split}
\widetilde{C}^{\Delta_i}_{3^0\rightarrow1^02^0}=&\mathcal{N}_{\Delta_1,0}\mathcal{N}_{\Delta_2,0}\mathcal{N}_{\Delta_3,0}\frac{g(2\pi)^7\Gamma^2[1-\Delta_1]\Gamma^2[1-\Delta_2]\Gamma[\Delta_1+\Delta_2-1]}{4\Gamma[\Delta_1]\Gamma[\Delta_2]\Gamma^2[2-\Delta_1-\Delta_2]}\delta(\Delta_1+\Delta_2+\Delta_3-2)\;.
\end{split}
\end{align}
Using \eqref{eq:phit} and the fact that
\begin{align}\label{eq:NNShadow=delta}
\mathcal{N}_{\Delta,0}\mathcal{N}_{2-\Delta,0}\int d^2z^{\prime}\frac{1}{|z_1-z^{\prime}|^{2\Delta}|z_2-z^{\prime}|^{2(2-\Delta)}}=\delta^{(2)}(z_1-z_2)\;,
\end{align}
we can obtain $\mathcal{A}^{\Delta_i}_{3^0\rightarrow1^02^0}$ from $\widetilde{\mathcal{A}}_{3^0\rightarrow 1^02^0}^{\Delta_i}$ through
\begin{align}
    \mathcal{A}^{\Delta_i}_{3^0\rightarrow1^02^0}=\mathcal{N}_{\Delta_3,0}\int d^2z_3^{\prime}\frac{1}{|z_3-z_3^{\prime}|^{2\Delta_3}}\widetilde{\mathcal{A}}_{3^0\rightarrow 1^02^0}^{\Delta_1,\Delta_2,2-\Delta_3}\;.
\end{align}
Substituting \eqref{eq:ShadowAmplitudeScalar} into the above equality and using \eqref{eq:triangle}, we find that $\mathcal{A}^{\Delta_i}_{1\rightarrow2}$ takes the form as a standard CFT three-point function with coefficient $C^{\Delta_i}_{3^0\rightarrow1^02^0}$ given by
\begin{align}
\begin{split}
   C^{\Delta_i}_{3^0\rightarrow1^02^0}=&(2\pi)^4\pi^2g\mathcal{N}_{\Delta_1,0}\mathcal{N}_{\Delta_2,0}\delta(\Delta_1+\Delta_2-\Delta_3)\\
   &\times\frac{\Gamma[1-\Delta_1]^2\Gamma[1-\Delta_2]^2\Gamma[\Delta_1+\Delta_2-1]\Gamma[2-\Delta_3]\Gamma[\frac{\Delta_{12}+\Delta_3}{2}]\Gamma[\frac{-\Delta_{12}+\Delta_3}{2}]}{\Gamma[\Delta_1]\Gamma[\Delta_2]\Gamma[\Delta_3-1]\Gamma[2-\Delta_1-\Delta_2]^2\Gamma[\frac{2+\Delta_1-\Delta_2-\Delta_3}{2}]\Gamma[\frac{2-\Delta_1+\Delta_2-\Delta_3}{2}]}\;.
\end{split}   
\end{align}
Setting $\Delta_3=\Delta_1+\Delta_2$ then reproduces \eqref{eq:ScalarTwoOutgoing}.

To compute the shadow celestial amplitude $\widetilde{\mathcal{A}}^{\Delta_i}_{1^0\rightarrow2^+_{\text{soft}}3^0}$ involving one incoming massless scalar, one outgoing photon and one outgoing massless scalar, we use \eqref{eq:PhotonPositionSpcae} to write $\widetilde{\mathcal{A}}^{\Delta_i}_{1^0\rightarrow2^+_{\text{soft}}3^0}$ as
\begin{align}
\begin{split}
\widetilde{\mathcal{A}}^{\Delta_i}_{1^0\rightarrow2^+_{\text{soft}}3^0}=&ie\int d^4X\;V^{\text{log},-}_{2,\mu z}\left[\partial^{\mu}\widetilde{\phi}_{\Delta_1}^{+}\widetilde{\phi}^{-}_{\Delta_3}- \widetilde{\phi}_{\Delta_1}^{+}\partial^{\mu}\widetilde{\phi}^{-}_{\Delta_3}\right]\;.
\end{split}
\end{align}
Following the steps in Section \ref{sec:PhotonScalarScalar}, $\widetilde{\mathcal{A}}^{\Delta_i}_{2\rightarrow1,z}$ can be rewritten as
\begin{align}
\begin{split}
\widetilde{\mathcal{A}}^{\Delta_i}_{1^0\rightarrow2^+_{\text{soft}}3^0}=&-2ie(\Delta_1-\Delta_3)\lim_{\Delta_2^{\prime}\rightarrow0}\frac{1}{N^-_{2-\Delta_2^{\prime}}\Delta_2^{\prime}}\partial_{z_2}\widetilde{\mathcal{A}}^{\Delta_3\Delta_2^{\prime}\Delta_1}_{1^0\rightarrow2^03^0}\;,
\end{split}
\end{align}
where $\widetilde{\mathcal{A}}^{\Delta_3\Delta_2^{\prime}\Delta_1}_{1^0\rightarrow2^03^0}$ is the scalar shadow celestial amplitude, obtained by switching $1$ and $3$ in \eqref{eq:ShadowAmplitudeScalar}. We note that the $\Delta_2^{\prime}\rightarrow0$ in the above equality dose not cause any subtlety since the delta-function in \eqref{eq:ShadowAmplitudeScalar} does not kill any $|z_{ij}|$ term. Using \eqref{eq:ShadowAmplitudeScalar}, we find that $\widetilde{\mathcal{A}}^{\Delta_i}_{1^0\rightarrow2^+_{\text{soft}}3^0}$ takes the standard form of conformal three-point function with conformal weights $(\Delta_1/2,\Delta_1/2)$, $(1,0)$ and $(\Delta_3/2,\Delta_3/2)$. The coefficient $\widetilde{C}^{\Delta_i}_{1^0\rightarrow2^+_{\text{soft}}3^0}$ is given by
\begin{align}\label{eq:ShadowPhotonScalarScalar}
\begin{split}
\widetilde{C}^{\Delta_i}_{1^0\rightarrow2^+_{\text{soft}}3^0}=&i(2\pi)^4e(\Delta_1-1)\delta(\Delta_1+\Delta_3-2)\;.
\end{split}
\end{align}
We can obtain $\mathcal{A}^{\Delta_i}_{1^0\rightarrow2^+_{\text{soft}}3^0}$ from $\widetilde{\mathcal{A}}_{1^0\rightarrow2^+_{\text{soft}}3^0}^{\Delta_i}$ through
\begin{align}
\mathcal{A}^{\Delta_i}_{1^0\rightarrow2^+_{\text{soft}}3^0}=\mathcal{N}_{\Delta_1,0}\int d^2z_1^{\prime}\frac{1}{|z_1-z_1^{\prime}|^{2\Delta_1}}\widetilde{\mathcal{A}}_{1^0\rightarrow2^+_{\text{soft}}3^0}^{2-\Delta_1,\Delta_2,\Delta_3}\;.
\end{align}
Substituting \eqref{eq:ShadowPhotonScalarScalar} into the above equality and restricting the prefactors to the support at  $\Delta_3=\Delta_1$ then reproduces \eqref{eq:PhotonScalarScalar}.

\bibliographystyle{ssg}
\bibliography{draft2}

\begingroup\raggedright\begin{thebibliography}{10}

\bibitem{Pasterski:2016qvg}
S.~Pasterski, S.-H. Shao, and A.~Strominger, ``{Flat Space Amplitudes and
  Conformal Symmetry of the Celestial Sphere},''
  \href{https://doi.org/10.1103/PhysRevD.96.065026}{{\it Phys. Rev. D}} {\bf
  96} (2017), no.~6 065026, \href{http://arxiv.org/abs/1701.00049}{{\tt
  1701.00049}}.

\bibitem{Strominger:2017zoo}
A.~Strominger, ``{Lectures on the Infrared Structure of Gravity and Gauge
  Theory},'' \href{http://arxiv.org/abs/1703.05448}{{\tt 1703.05448}}.

\bibitem{Raclariu:2021zjz}
A.-M. Raclariu, ``{Lectures on Celestial Holography},''
  \href{http://arxiv.org/abs/2107.02075}{{\tt 2107.02075}}.

\bibitem{Pasterski:2021rjz}
S.~Pasterski, ``{Lectures on celestial amplitudes},''
  \href{https://doi.org/10.1140/epjc/s10052-021-09846-7}{{\it Eur. Phys. J. C}}
  {\bf 81} (2021), no.~12 1062, \href{http://arxiv.org/abs/2108.04801}{{\tt
  2108.04801}}.

\bibitem{Pasterski:2021raf}
S.~Pasterski, M.~Pate, and A.-M. Raclariu, ``{Celestial Holography},'' in {\em
  {2022 Snowmass Summer Study}}, 11, 2021.
\newblock \href{http://arxiv.org/abs/2111.11392}{{\tt 2111.11392}}.

\bibitem{Weinberg:1965nx}
S.~Weinberg, ``{Infrared photons and gravitons},''
  \href{https://doi.org/10.1103/PhysRev.140.B516}{{\it Phys. Rev.}} {\bf 140}
  (1965) B516--B524.

\bibitem{He:2014laa}
T.~He, V.~Lysov, P.~Mitra, and A.~Strominger, ``{BMS supertranslations and
  Weinberg\textquoteright{}s soft graviton theorem},''
  \href{https://doi.org/10.1007/JHEP05(2015)151}{{\it JHEP}} {\bf 05} (2015)
  151, \href{http://arxiv.org/abs/1401.7026}{{\tt 1401.7026}}.

\bibitem{Kapec:2016jld}
D.~Kapec, P.~Mitra, A.-M. Raclariu, and A.~Strominger, ``{2D Stress Tensor for
  4D Gravity},'' \href{https://doi.org/10.1103/PhysRevLett.119.121601}{{\it
  Phys. Rev. Lett.}} {\bf 119} (2017), no.~12 121601,
  \href{http://arxiv.org/abs/1609.00282}{{\tt 1609.00282}}.

\bibitem{Donnay:2018neh}
L.~Donnay, A.~Puhm, and A.~Strominger, ``{Conformally Soft Photons and
  Gravitons},'' \href{https://doi.org/10.1007/JHEP01(2019)184}{{\it JHEP}} {\bf
  01} (2019) 184, \href{http://arxiv.org/abs/1810.05219}{{\tt 1810.05219}}.

\bibitem{Stieberger:2018onx}
S.~Stieberger and T.~R. Taylor, ``{Symmetries of Celestial Amplitudes},''
  \href{https://doi.org/10.1016/j.physletb.2019.03.063}{{\it Phys. Lett. B}}
  {\bf 793} (2019) 141--143, \href{http://arxiv.org/abs/1812.01080}{{\tt
  1812.01080}}.

\bibitem{Fan:2019emx}
W.~Fan, A.~Fotopoulos, and T.~R. Taylor, ``{Soft Limits of Yang-Mills
  Amplitudes and Conformal Correlators},''
  \href{https://doi.org/10.1007/JHEP05(2019)121}{{\it JHEP}} {\bf 05} (2019)
  121, \href{http://arxiv.org/abs/1903.01676}{{\tt 1903.01676}}.

\bibitem{Pate:2019mfs}
M.~Pate, A.-M. Raclariu, and A.~Strominger, ``{Conformally Soft Theorem in
  Gauge Theory},'' \href{https://doi.org/10.1103/PhysRevD.100.085017}{{\it
  Phys. Rev. D}} {\bf 100} (2019), no.~8 085017,
  \href{http://arxiv.org/abs/1904.10831}{{\tt 1904.10831}}.

\bibitem{Adamo:2019ipt}
T.~Adamo, L.~Mason, and A.~Sharma, ``{Celestial amplitudes and conformal soft
  theorems},'' \href{https://doi.org/10.1088/1361-6382/ab42ce}{{\it Class.
  Quant. Grav.}} {\bf 36} (2019), no.~20 205018,
  \href{http://arxiv.org/abs/1905.09224}{{\tt 1905.09224}}.

\bibitem{Puhm:2019zbl}
A.~Puhm, ``{Conformally Soft Theorem in Gravity},''
  \href{https://doi.org/10.1007/JHEP09(2020)130}{{\it JHEP}} {\bf 09} (2020)
  130, \href{http://arxiv.org/abs/1905.09799}{{\tt 1905.09799}}.

\bibitem{Guevara:2019ypd}
A.~Guevara, ``{Notes on Conformal Soft Theorems and Recursion Relations in
  Gravity},'' \href{http://arxiv.org/abs/1906.07810}{{\tt 1906.07810}}.

\bibitem{Fotopoulos:2019tpe}
A.~Fotopoulos and T.~R. Taylor, ``{Primary Fields in Celestial CFT},''
  \href{https://doi.org/10.1007/JHEP10(2019)167}{{\it JHEP}} {\bf 10} (2019)
  167, \href{http://arxiv.org/abs/1906.10149}{{\tt 1906.10149}}.

\bibitem{Strominger:2021lvk}
A.~Strominger, ``{w(1+infinity) and the Celestial Sphere},''
  \href{http://arxiv.org/abs/2105.14346}{{\tt 2105.14346}}.

\bibitem{Ball:2021tmb}
A.~Ball, S.~A. Narayanan, J.~Salzer, and A.~Strominger, ``{Perturbatively exact
  w$_{1+\infty}$ asymptotic symmetry of quantum self-dual gravity},''
  \href{https://doi.org/10.1007/JHEP01(2022)114}{{\it JHEP}} {\bf 01} (2022)
  114, \href{http://arxiv.org/abs/2111.10392}{{\tt 2111.10392}}.

\bibitem{Strominger:2021mtt}
A.~Strominger, ``{$w_{1+\infty}$ Algebra and the Celestial Sphere: Infinite
  Towers of Soft Graviton, Photon, and Gluon Symmetries},''
  \href{https://doi.org/10.1103/PhysRevLett.127.221601}{{\it Phys. Rev. Lett.}}
  {\bf 127} (2021), no.~22 221601.

\bibitem{Pasterski:2017ylz}
S.~Pasterski, S.-H. Shao, and A.~Strominger, ``{Gluon Amplitudes as 2d
  Conformal Correlators},''
  \href{https://doi.org/10.1103/PhysRevD.96.085006}{{\it Phys. Rev. D}} {\bf
  96} (2017), no.~8 085006, \href{http://arxiv.org/abs/1706.03917}{{\tt
  1706.03917}}.

\bibitem{Stieberger:2018edy}
S.~Stieberger and T.~R. Taylor, ``{Strings on Celestial Sphere},''
  \href{https://doi.org/10.1016/j.nuclphysb.2018.08.019}{{\it Nucl. Phys. B}}
  {\bf 935} (2018) 388--411, \href{http://arxiv.org/abs/1806.05688}{{\tt
  1806.05688}}.

\bibitem{Pate:2019lpp}
M.~Pate, A.-M. Raclariu, A.~Strominger, and E.~Y. Yuan, ``{Celestial operator
  products of gluons and gravitons},''
  \href{https://doi.org/10.1142/S0129055X21400031}{{\it Rev. Math. Phys.}} {\bf
  33} (2021), no.~09 2140003, \href{http://arxiv.org/abs/1910.07424}{{\tt
  1910.07424}}.

\bibitem{Banerjee:2020kaa}
S.~Banerjee, S.~Ghosh, and R.~Gonzo, ``{BMS symmetry of celestial OPE},''
  \href{https://doi.org/10.1007/JHEP04(2020)130}{{\it JHEP}} {\bf 04} (2020)
  130, \href{http://arxiv.org/abs/2002.00975}{{\tt 2002.00975}}.

\bibitem{Jiang:2021xzy}
H.~Jiang, ``{Celestial superamplitude in $ \mathcal{N} $ = 4 SYM theory},''
  \href{https://doi.org/10.1007/JHEP08(2021)031}{{\it JHEP}} {\bf 08} (2021)
  031, \href{http://arxiv.org/abs/2105.10269}{{\tt 2105.10269}}.

\bibitem{Brandhuber:2021nez}
A.~Brandhuber, G.~R. Brown, J.~Gowdy, B.~Spence, and G.~Travaglini,
  ``{Celestial superamplitudes},''
  \href{https://doi.org/10.1103/PhysRevD.104.045016}{{\it Phys. Rev. D}} {\bf
  104} (2021), no.~4 045016, \href{http://arxiv.org/abs/2105.10263}{{\tt
  2105.10263}}.

\bibitem{Sharma:2021gcz}
A.~Sharma, ``{Ambidextrous light transforms for celestial amplitudes},''
  \href{https://doi.org/10.1007/JHEP01(2022)031}{{\it JHEP}} {\bf 01} (2022)
  031, \href{http://arxiv.org/abs/2107.06250}{{\tt 2107.06250}}.

\bibitem{Crawley:2021ivb}
E.~Crawley, N.~Miller, S.~A. Narayanan, and A.~Strominger, ``{State-operator
  correspondence in celestial conformal field theory},''
  \href{https://doi.org/10.1007/JHEP09(2021)132}{{\it JHEP}} {\bf 09} (2021)
  132, \href{http://arxiv.org/abs/2105.00331}{{\tt 2105.00331}}.

\bibitem{Pasterski:2017kqt}
S.~Pasterski and S.-H. Shao, ``{Conformal basis for flat space amplitudes},''
  \href{https://doi.org/10.1103/PhysRevD.96.065022}{{\it Phys. Rev. D}} {\bf
  96} (2017), no.~6 065022, \href{http://arxiv.org/abs/1705.01027}{{\tt
  1705.01027}}.

\bibitem{Law:2020tsg}
Y.~T.~A. Law and M.~Zlotnikov, ``{Massive Spinning Bosons on the Celestial
  Sphere},'' \href{https://doi.org/10.1007/JHEP06(2020)079}{{\it JHEP}} {\bf
  06} (2020) 079, \href{http://arxiv.org/abs/2004.04309}{{\tt 2004.04309}}.

\bibitem{Caron-Huot:2022eqs}
S.~Caron-Huot, M.~Kologlu, P.~Kravchuk, D.~Meltzer, and D.~Simmons-Duffin,
  ``{Detectors in weakly-coupled field theories},''
  \href{http://arxiv.org/abs/2209.00008}{{\tt 2209.00008}}.

\bibitem{Chang:2022jut}
C.-M. Chang, W.~Cui, W.-J. Ma, H.~Shu, and H.~Zou, ``{Shadow Celestial
  Amplitude},'' \href{http://arxiv.org/abs/2210.04725}{{\tt 2210.04725}}.

\bibitem{Strominger:2013jfa}
A.~Strominger, ``{On BMS Invariance of Gravitational Scattering},''
  \href{https://doi.org/10.1007/JHEP07(2014)152}{{\it JHEP}} {\bf 07} (2014)
  152, \href{http://arxiv.org/abs/1312.2229}{{\tt 1312.2229}}.

\bibitem{Barnich:2013axa}
G.~Barnich and C.~Troessaert, ``{Comments on holographic current algebras and
  asymptotically flat four dimensional spacetimes at null infinity},''
  \href{https://doi.org/10.1007/JHEP11(2013)003}{{\it JHEP}} {\bf 11} (2013)
  003, \href{http://arxiv.org/abs/1309.0794}{{\tt 1309.0794}}.

\bibitem{Chang:2021wvv}
C.-M. Chang, Y.-t. Huang, Z.-X. Huang, and W.~Li, ``{Bulk locality from the
  celestial amplitude},''
  \href{https://doi.org/10.21468/SciPostPhys.12.5.176}{{\it SciPost Phys.}}
  {\bf 12} (2022), no.~5 176, \href{http://arxiv.org/abs/2106.11948}{{\tt
  2106.11948}}.

\bibitem{Cachazo:2014fwa}
F.~Cachazo and A.~Strominger, ``{Evidence for a New Soft Graviton Theorem},''
  \href{http://arxiv.org/abs/1404.4091}{{\tt 1404.4091}}.

\bibitem{Kapec:2014opa}
D.~Kapec, V.~Lysov, S.~Pasterski, and A.~Strominger, ``{Semiclassical Virasoro
  symmetry of the quantum gravity $ \mathcal{S}$-matrix},''
  \href{https://doi.org/10.1007/JHEP08(2014)058}{{\it JHEP}} {\bf 08} (2014)
  058, \href{http://arxiv.org/abs/1406.3312}{{\tt 1406.3312}}.

\bibitem{Fotopoulos:2020bqj}
A.~Fotopoulos, S.~Stieberger, T.~R. Taylor, and B.~Zhu, ``{Extended Super BMS
  Algebra of Celestial CFT},''
  \href{https://doi.org/10.1007/JHEP09(2020)198}{{\it JHEP}} {\bf 09} (2020)
  198, \href{http://arxiv.org/abs/2007.03785}{{\tt 2007.03785}}.

\bibitem{Ball:2022bgg}
A.~Ball, ``{Celestial Locality and the Jacobi Identity},''
  \href{http://arxiv.org/abs/2211.09151}{{\tt 2211.09151}}.

\bibitem{Banerjee:2022wht}
S.~Banerjee and S.~Pasterski, ``{Revisiting the Shadow Stress Tensor in
  Celestial CFT},'' \href{http://arxiv.org/abs/2212.00257}{{\tt 2212.00257}}.

\bibitem{Jorge-Diaz:2022dmy}
C.~Jorge-Diaz, S.~Pasterski, and A.~Sharma, ``{Celestial amplitudes in an
  ambidextrous basis},'' \href{http://arxiv.org/abs/2212.00962}{{\tt
  2212.00962}}.

\bibitem{Brown:2022miw}
G.~R. Brown, J.~Gowdy, and B.~Spence, ``{Celestial Twistor Amplitudes},''
  \href{http://arxiv.org/abs/2212.01327}{{\tt 2212.01327}}.

\bibitem{Weinberg:2010fx}
S.~Weinberg, ``{Six-dimensional Methods for Four-dimensional Conformal Field
  Theories},'' \href{https://doi.org/10.1103/PhysRevD.82.045031}{{\it Phys.
  Rev. D}} {\bf 82} (2010) 045031, \href{http://arxiv.org/abs/1006.3480}{{\tt
  1006.3480}}.

\bibitem{Weinberg:2012mz}
S.~Weinberg, ``{Six-dimensional Methods for Four-dimensional Conformal Field
  Theories II: Irreducible Fields},''
  \href{https://doi.org/10.1103/PhysRevD.86.085013}{{\it Phys. Rev. D}} {\bf
  86} (2012) 085013, \href{http://arxiv.org/abs/1209.4659}{{\tt 1209.4659}}.

\bibitem{Rychkov:2016iqz}
S.~Rychkov, {\em {EPFL Lectures on Conformal Field Theory in D\ensuremath{>}= 3
  Dimensions}}.
\newblock SpringerBriefs in Physics. 1, 2016.

\bibitem{Fortin:2019dnq}
J.-F. Fortin and W.~Skiba, ``{New methods for conformal correlation
  functions},'' \href{https://doi.org/10.1007/JHEP06(2020)028}{{\it JHEP}} {\bf
  06} (2020) 028, \href{http://arxiv.org/abs/1905.00434}{{\tt 1905.00434}}.

\bibitem{Skiba:2019cmf}
W.~Skiba and J.-F. Fortin, ``{A Recipe for Conformal Blocks},''
  \href{https://doi.org/10.31526/lhep.2022.293}{{\it LHEP}} {\bf 2022} (2022)
  293, \href{http://arxiv.org/abs/1905.00036}{{\tt 1905.00036}}.

\bibitem{Atanasov:2021cje}
A.~Atanasov, W.~Melton, A.-M. Raclariu, and A.~Strominger, ``{Conformal block
  expansion in celestial CFT},''
  \href{https://doi.org/10.1103/PhysRevD.104.126033}{{\it Phys. Rev. D}} {\bf
  104} (2021), no.~12 126033, \href{http://arxiv.org/abs/2104.13432}{{\tt
  2104.13432}}.

\bibitem{Dolan:2011dv}
F.~A. Dolan and H.~Osborn, ``{Conformal Partial Waves: Further Mathematical
  Results},'' \href{http://arxiv.org/abs/1108.6194}{{\tt 1108.6194}}.

\end{thebibliography}\endgroup

\end{document}